\documentclass[preprint2]{aastex}

\usepackage{graphicx}% Include figure files

\newcommand{\ran}{($\alpha$,n)}

\newcommand{\rgn}{($\gamma$,n)}
\newcommand{\rng}{(n,$\gamma$)}

\begin{document}

\title{Opportunities to constrain astrophysical reaction rates for the $s$ process
through determination of the ground state cross sections}

\author{T. Rauscher\altaffilmark{1}, P. Mohr\altaffilmark{2,5}, I. Dillmann\altaffilmark{3,6}, and R. Plag\altaffilmark{4,6}}

\altaffiltext{1}{Department of Physics, University of Basel, CH-4056 Basel, Switzerland}
\altaffiltext{2}{Diakonie-Klinikum, D-74523 Schw\"abisch Hall, Germany}
\altaffiltext{3}{II.\ Physikalisches Institut, Justus-Liebig-Universit\"at, D-35392 Giessen, Germany}
\altaffiltext{4}{Institut f\"ur Angewandte Physik, Goethe-Universit\"at, D-60438 Frankfurt am Main, Germany}
\altaffiltext{5}{Institute of Nuclear Research (ATOMKI), H-4001 Debrecen, Hungary}
\altaffiltext{6}{GSI Helmholtzzentrum f\"ur Schwerionenforschung GmbH, D-64291 Darmstadt, Germany}

%\date{\today}

\begin{abstract}
Modern models of $s$-process nucleosynthesis in stars require stellar reaction rates with high precision. Most of the
neutron capture cross sections in the $s$ process have been measured and for an increasing number of reactions the required precision
is achieved. This does not necessarily mean, however, that the stellar rates are constrained equally well because only capture on the ground state of
a target is measured in the laboratory. Captures on excited states can considerably contribute to stellar rates already at typical $s$-process
temperatures. We show that the ground state contribution $X$ to a stellar rate is the relevant measure to identify reactions which are or
could be well constrained by experiments and apply it to \rng\ reactions in the $s$ process. It is further shown that the
maximally possible reduction in uncertainty of a rate through determination of the g.s.\ cross section is directly given by $X$.
An error analysis of $X$ is presented and it is found that $X$ is a robust measure with overall small uncertainties.
Several specific examples (neutron capture on $^{79}$Se, $^{95}$Zr,
$^{121}$Sn, $^{187}$Os, and $^{193}$Pt) are discussed in detail. The ground state contributions for a set of 411 neutron capture
reactions around the $s$-process path are presented in a table. This allows to identify reactions which may be better constrained by experiments and such which cannot be constrained by only measuring ground state cross sections (and thus require supplementary studies). General trends and implications are discussed.
\end{abstract}

%\pacs{26.20.-f, 26.20.Kn, 25.40.Lw}
% 26.20.-f 	Hydrostatic stellar nucleosynthesis
% 26.20.Kn 	s-process
% 25.40.Lw 	Radiative capture

%\maketitle

\section{Introduction}
The astrophysical $s$ process is the probably best understood nucleosynthesis
process, contributing about half of the natural nuclides beyond Fe
\citep{Kae06,Str06,Kae11}. The main component of the $s$ process is produced in
He-shell flashes of Asymptotic Giant Branch (AGB) stars. A second, weak
component is from massive stars, additionally contributing to the nuclei in
the mass range $A\lesssim 90$. The $s$ process proceeds by sequences of
neutron capture reactions to build neutron-richer isotopes within an isotopic
chain. As soon as an unstable isotope is reached, $\beta^-$-decay moves the
synthesis path to the next element with $Z+1$. In some cases neutron capture and
$\beta^-$-decay compete because of comparable reaction rates.  This
establishes branching points in the $s$-process path. The resulting abundances
depend sensitively on the nuclear properties of these branching points. This
fact may be used either to derive astrophysical parameters, such as
temperature and neutron density during the $s$ process, or to test stellar
$s$-process models which provide these parameters.

One prerequisite for the success of nucleosynthesis calculations for the $s$ process
is the availability of excellent experimental neutron capture data \citep[see compilations in][]{Bao00,KADONISV0,KADONIS}. As the $s$-process path
proceeds along the valley of stability, most of the nuclei (except the
unstable branching nuclei) can be used as stable targets. Typical $s$-process
temperatures are of the order of $kT \approx 5-30$\,keV for the main component
in the He-shell flashes of AGB stars, where the main neutron source $^{13}$C\ran $^{16}$O operates at the lower end of the
temperature range and the second neutron producing reaction $^{22}$Ne\ran $^{25}$Mg at the upper end \citep{Kae06}.
The latter reaction at slightly higher temperatures of up to about 80 keV also provides the neutrons for the weak $s$ process
component in massive stars \citep{rhhw02}. Experiments on stable targets are feasible at these
energies, with typical cross sections ranging from milli-barns up to a few
barns. However, many reactions on stable targets are still not measured in the full energy range important in the $s$ process.
In the near future, also cross section measurements on unstable nuclei in the branchings will become possible.

The central quantities in all $s$-process calculations are the astrophysical reaction rates. These have to be determined by
theoretical predictions in the absence of experimental data in the relevant energy range. Due to the impact of thermal excitation
of nuclear states above the ground state in the stellar plasma, however, a theoretical contribution may remain even when the
ground state cross sections are measured. Although the $s$ process involves comparatively low plasma temperatures,
transitions from low-lying excited states may contribute considerably to the stellar rate.

We studied the contribution of ground state transitions to stellar neutron capture rates for the $s$ process with the aim to
identify cases for which the astrophysical reaction rate is mainly given by the ground state rate and thus can be well constrained by
the measurement of ground state cross sections. At the same time this allows to pinpoint reactions for which the uncertainty in the
astrophysical reaction rate will not be removed by the determination of ground state cross sections. This is also interesting to
$s$-process modelers who want to estimate the uncertainties in the rates they are using. We also show a general method
to determine how the uncertainties in the stellar rates are modified when better constraining one or more transitions from a
ground or excited state, not limited to neutron captures in the $s$ process.

\section{Stellar rates}
\subsection{Definition}

Nuclear reactions occurring in an astrophysical plasma are described by
astrophysical reaction rates; e.g., for neutron capture the rate is defined by
\citep{fow74,tommyreview}
\begin{equation}
r^*=n_\mathrm{n} \, n_\mathrm{target} \, \langle \sigma v \rangle ^* =
n_\mathrm{n} \, n_\mathrm{target} \, R^*\quad,
\label{eq:rate_star}
\end{equation}
with $n_\mathrm{n}$, $n_\mathrm{target}$ being number densities of the
neutrons and a species of target nuclei, and the stellar rate factor (also
called reactivity) $R^* = \langle \sigma v \rangle^*$. The relative velocities
$v$ between the neutrons and the
target nuclei are given by the Maxwell-Boltzmann (MB) distribution
$\Phi_{\rm{MB}}(v,T)$ at temperature $T$. It can also be expressed as function
of relative interaction energy $\Phi(E,T)$.

In a stellar plasma, nuclei are in thermal equilibrium with the plasma. This
also implies that internal states can be excited and the nuclei do not have to
be in the ground state (g.s.) necessarily, depending on the temperature. In
most cases, the excited states within a nucleus are also in thermal
equilibrium because the equilibration timescale is short \citep{Gin09}. There
are few cases with isomeric or long-lived states that may not be thermalized at
$s$-process temperatures, such as $^{176}$Lu \citep{Heil08,Mohr09,Gin09} or
$^{180}$Ta \citep{Bel99,Mohr07}. Assuming thermal equilibrium, the total
stellar reactivity $R^*$ is \citep[see, e.g.,][]{WF80}
\begin{equation}
R^*(T) = w_0 R_0 + w_1 R_1 + w_2 R_2 + \dots \quad
\label{eq:stellarrate}
\end{equation}
with the reactivities of the $i$-th state in the target
\begin{equation}
R_i(T) = \int_0^\infty \sigma_i(E_i) \, \Phi(E_i,T) \, dE_i \quad .
\label{eq:reactivity}
\end{equation}
The cross section $\sigma_i$ is the total neutron capture cross section of the
$i$-th state, i.e.\ summed over all final states $j$ in the
residual ($A+1$) nucleus: $\sigma_i = \sum_j \sigma^{i\rightarrow j}$.
The relative energy $E_i$ between projectile and target is always measured in the system of the respective target state
(i.e., g.s.\ or excited state) and
therefore the integration always starts at zero \citep{fow74,tommyreview}.
Thus, the full MB energy distributed neutrons are also acting on
nuclei which are in excited states, not only on nuclei in the g.s., and the
reaction rate in Equation (\ref{eq:rate_star}) has to be defined
appropriately by the sum in Equation (\ref{eq:reactivity}).

The normalized weights $w_i$ in Equation (\ref{eq:stellarrate}) of the excited
state contributions are given by
\begin{equation}
w_i=\frac{P_i}{\sum_i P_i}=\frac{P_i}{G} \quad,
\end{equation}
with the partition function $G=\sum_i P_i$ or, normalized to the g.s.,
$G_0=G/(2J_0+1)$. The population factor $P_i$ of the excited state $i$ with
excitation energy $E_i$ and spin $J_i$ is given by the usual Boltzmann factor
\begin{equation}
\label{eq:boltzmannfactor}
P_i=(2J_i+1)\exp\left(-\frac{E_i}{kT}\right) \quad.
\end{equation}
Normalized partition functions $G_0(T)$ are usually given along with
reaction rate predictions and can be found, e.g., in \citet{holadndt,wooadndt,Rau00}.

Often, the integral in Equation (\ref{eq:reactivity}) is divided by the mean thermal
velocity $v_{i,T}$, leading to the so-called Maxwellian averaged cross section
(MACS). The MACS for the g.s.\ ($i=0$) can be derived from a
measurement of the neutron capture cross sections over a properly chosen energy
interval. This MACS for the g.s.\ and its uncertainty is provided in the
compilations \citet{Bao00,KADONISV0,KADONIS}.

Laboratory experiments are performed on nuclei in the g.s.\ (with the
exception of $^{180}$Ta which is the only naturally occuring isomer and has a
short-lived ground state). Thus, experiments only determine the neutron
capture cross section of the g.s., i.e., $\sigma_0$. Therefore, only the
g.s.\ reactivity $R_0$ (or g.s.\ MACS) can be derived instead of the desired
astrophysical reactivity $R^*$. To assess by how much
the g.s.\ rate differs from the astrophysical rate, usually the so-called
``stellar enhancement factor'' (SEF)
\begin{equation}
f_\mathrm{SEF} = \frac{R^*(T)}{R_0}
\label{eq:SEF}
\end{equation}
is used. It is found that the SEF remains relatively close to unity within about $20-30\,\%$ for
most nuclei in the $s$-process path up to temperatures of $kT \approx
100$\,keV \citep{Bao00,KADONISV0,KADONIS}. As shown below, however, this does not necessarily imply a small contribution
of transitions from excited states or small remaining uncertainty once $R_0$ has been well determined.

\subsection{Excited state contributions}
\label{sec:excitedstates}

It can be shown that the weighted
sum of integrals in Equation (\ref{eq:stellarrate}) can be
transformed to an integral over a \textit{single} MB distribution
\citep{fow74,holadndt,tommyreview}
\begin{equation}
R^*= \frac{1}{G_0(T)} \int_0^\infty \sigma^\mathrm{eff}(E_0) \,
\Phi(E_0,T) \, dE_0 \quad,
\label{eq:effrate}
\end{equation}
thereby avoiding the different energy scales found in Equation (\ref{eq:stellarrate}).
Following \citet{fow74,holadndt}, $\sigma^\mathrm{eff}$ is the so-called
effective cross section, summing over all final states \textit{and} initial excited states up to the interaction energy $E$
\begin{equation}
\sigma^\mathrm{eff}(E)=\sum_i \sum_j \frac{2J_i+1}{2J_0+1} \frac{E-E_i}{E}
\sigma^{i \rightarrow j}(E-E_i) \quad.
\end{equation}

This transformation proves convenient in determining the actual weights with which transitions from excited states $i$ are
contributing to the stellar rate. These are not the $P_i$ as defined in Equation (\ref{eq:boltzmannfactor}) because of the different MB
distributions appearing in Equation (\ref{eq:stellarrate}). They rather are \citep{tommyreview}
\begin{equation}
\label{eq:weights}
\mathcal{W}_i=(2J_i+1)\frac{E-E_i}{E}=(2J_i+1)\left(1-\frac{E_i}{E}\right)
\end{equation}
and therefore \textit{linearly} declining with increasing excitation energy $E_i$. The energies $E$ appearing
in Equation (\ref{eq:weights}) are given by the range of energies significantly contributing to the integral in Equation (\ref{eq:effrate}).
Therefore the maximum reference energy $E$ appearing is the upper end of the relevant energy window (Gamow window). The relevant
energy window for neutron captures is located close to the peak of the MB distribution and its width is comparable to the width
of the MB distribution \citep{iliadisbook,energywindows}. As can be seen in \citet{iliadisbook,energywindows},
the upper edge of the effective energy window for neutrons is located not far above the energy $E=kT$, leading to
$\mathcal{W}_i \approx (2J_i+1)\left(1-E_i/(kT)\right)$.

The above shows that for the $s$-process temperature of $kT=30$ keV, excited states up to $\approx 30-40$ keV excitation energy can be important in the target nucleus
due to the linearly decreasing weight, for a temperature of $kT=80$ keV excited states even up to $\approx 80-100$ keV have to be considered. Obviously,
$\sigma^{i \rightarrow j}$ contains the information on how strong the contribution from the excited state is. But $\mathcal{W}$ allows us to estimate whether we have to bother with calculating $\sigma^{i \rightarrow j}$ at all.

Despite the fact that many nuclei in the $s$-process path show excited states below 100 keV, especially the heavier nuclides, the SEFs nevertheless remain close to unity for most cases. Clearly, this implies that the (predicted) stellar reactivity is close to the one for the g.s.\ because either the $w_i$ ($\mathcal{W}_i$) or the $\sigma^{i \rightarrow j}$ are negligible for $i>0$, or they conspire to yield a value close to the one obtained for the g.s.\ transitions alone. The SEF does not allow to distinguish between those cases. For illustration, we examine a nucleus with only two levels
characterized by excitation energies $E_0 = 0$ and $E_1$, spins
$J_0$ and $J_1$, and thermal occupation weights $w_0$ and $w_1$.
Let us consider two different cases in this simple model.
For case A we assume that 90\,\% of the target nuclei are in the g.s.
($w_0 = 0.9$) and 10\,\% are in the excited state ($w_1 = 0.1$). Let us
further assume that the cross section of the excited state $\sigma_1$ is
proportional to $\sigma_0$, but larger by a factor of two. This yields
\begin{equation}
f_\mathrm{SEF}^{\mathrm{A}} = \frac{0.9 R_0 + 0.1 R_1}{R_0} = 0.9 + 0.1 \times
2 = 1.1 \quad.
\label{eq:SEFA}
\end{equation}
For case B we assume that
$2/3$ of the target nuclei are in the first excited state and only 1/3 remain
in the g.s. This can be easily achieved for a small excitation energy
$E_1 \ll kT$ and $J_0 \ll J_1$. Further, we assume that the cross section
$\sigma_1$ is proportional to $\sigma_0$, but larger by 15\,\%. This results in
\begin{equation}
f_\mathrm{SEF}^{\mathrm{B}} = \frac{R_0/3 + 2R_1/3}{R_0} = 1/3 + 2/3 \times
1.15 = 1.1  \quad.
\label{eq:SEFB}
\end{equation}
Although the SEFs are identical in both cases and close to unity, the
essential difference is in
the contribution of the excited state, which is
small in case A but dominant in case B.

\subsection{Ground state contribution}
\label{sec:gscontrib}

Instead of using the SEF, the importance of excited states in the stellar rate factor $R^*$ can be better assessed by
the g.s.\ contribution $X$ to $R^*$ which is given by
\begin{equation}
X = \frac{w_0 R_0}{R^*} = \frac{w_0}{f_\mathrm{SEF}}
\label{eq:frac}
\end{equation}
for the two simple examples above.
The quantity $X$ is positive and assumes its maximal value of unity when only
the target g.s.\ is contributing to $R^*$.
It follows that in case A the g.s.\ contribution to the stellar rate factor
is $X = 0.9 / 1.1 \approx 0.82$, i.e.\ the stellar rate factor
is essentially defined by the dominating g.s.\
contribution. On the other hand, in case B we find a g.s.\
contribution of only $X = (1/3)/ 1.1 \approx 0.30$, i.e.\ about 70\,\% of the
stellar rate are made by the excited state. This shows that the excited state
contribution may exceed the g.s.\ contribution significantly, even when
$f_\mathrm{SEF} \approx 1$.

The above result for the g.s.\ contribution can easily be generalized
for many contributing thermally excited states.
Equation \ref{eq:effrate} clearly shows that $f_\mathrm{SEF}$ is \textit{not} constantly unity without the contribution of excited state transitions
but, rather, $f_\mathrm{SEF}=1/G_0$ in this case \citep{tommyreview}.
Therefore the g.s.\ contribution $X\leq 1$ in the general case is given by
\begin{equation}
X(T) = \frac{1}{f_\mathrm{SEF}(T) \, G_0(T)} = \frac{R_0}{R^* G_0} \quad.
\label{eq:X}
\end{equation}
The quantity $X$ will be included in future versions of the KADoNiS compilation \citep{KADONIS}.
Even when it is not directly given in a compilation, however, it can be easily computed from $G_0$ and either $f_\mathrm{SEF}$ or $R_0$, as shown above.
The SEF itself still retains
some interest in the sense that it specifies whether the stellar rate,
although incidentally, is similar to the g.s.\ rate.

\subsection{Uncertainty in $X$}
\label{sec:uncert}

It has to be kept in mind, however, that $X$ itself is a theoretical quantity
(like the SEF). The weighted uncertainties in the rates or cross sections of
g.s.\ and excited states determine the total uncertainty in $X$ and
$f_\mathrm{SEF}$. Entering this total uncertainty, however, are the uncertainties in the \textit{ratios}
of the rates of excited states and g.s.\ as well as the uncertainty in the ratios of the weights, i.e.\ of $G_0$.
This can be easily seen when looking at the reciprocal of $X$ (which must have the same relative uncertainty)
\begin{equation}
\label{eq:1x}
\frac{1}{X}=1+\frac{w_1}{w_0} \frac{R_1}{R_0}+\frac{w_2}{w_0} \frac{R_2}{R_0}+\frac{w_3}{w_0} \frac{R_3}{R_0}+\dots
\end{equation}
Similar considerations apply to the SEF.
For example, using the simple two-level system from Section \ref{sec:excitedstates}, we
may assign some uncertainty, say a factor of two, to the term $w_1R_1/(w_0R_0)$. This translates into uncertainties
$f_\mathrm{SEF}^{\mathrm{A}} = 1.1^{+0.20}_{-0.10}$,
$X^{\mathrm{A}}=0.82^{+0.08}_{-0.13}$, and
$f_\mathrm{SEF}^{\mathrm{B}} =
1.1^{+0.77}_{-0.38}$, $X^{\mathrm{B}}=0.30^{+0.16}_{-0.12}$.

In the following we study the influence of an uncertainty factor $u$ in the
calculations on the resulting ground state contribution $X$ and its
inverse $1/X$.
Only nuclei at or close to stability participate in the $s$ process. These nuclides have generally well determined level
schemes at low excitation energies and therefore $G_0$ and the weights $w_i$ are well determined. Thus, the uncertainties
in $X$ are limited to the uncertainties $u_i$ arising from the predicted ratios $R_i/R_0$,
\begin{equation}
\label{eq:1xuncert}
u_{1/X}=1+\frac{w_1}{w_0} u_{1}\frac{R_1}{R_0} +\frac{w_2}{w_0} u_2\frac{R_2}{R_0} +\frac{w_3}{w_0} u_3 \frac{R_3}{R_0} +\dots
\end{equation}
For the two-level system above we had assumed an uncertainty of a factor of two, i.e.\ $u_1=2$ and $u_1=1/2$, to arrive at the given total uncertainty in $X$.

Another convenient property of $X$ is that its uncertainty scales with $X$ itself. This can be seen when assuming
$\overline{u}=u_1=u_2=u_3=\dots$ For realistic cases in the $s$ process
this is a good approximation to the actual errors in the ratios because only the first few
terms in the sum will contribute to $X$ and the weighted average of the uncertainties $\overline{u}$ will be close to the values of
$u_1$ and/or $u_2$. The applicability of the assumption can be checked through the normalized partition function. The closer $G_0$ is
to unity, the better the above assumption holds.

\begin{figure}
\includegraphics[angle=-90,width=1\columnwidth]{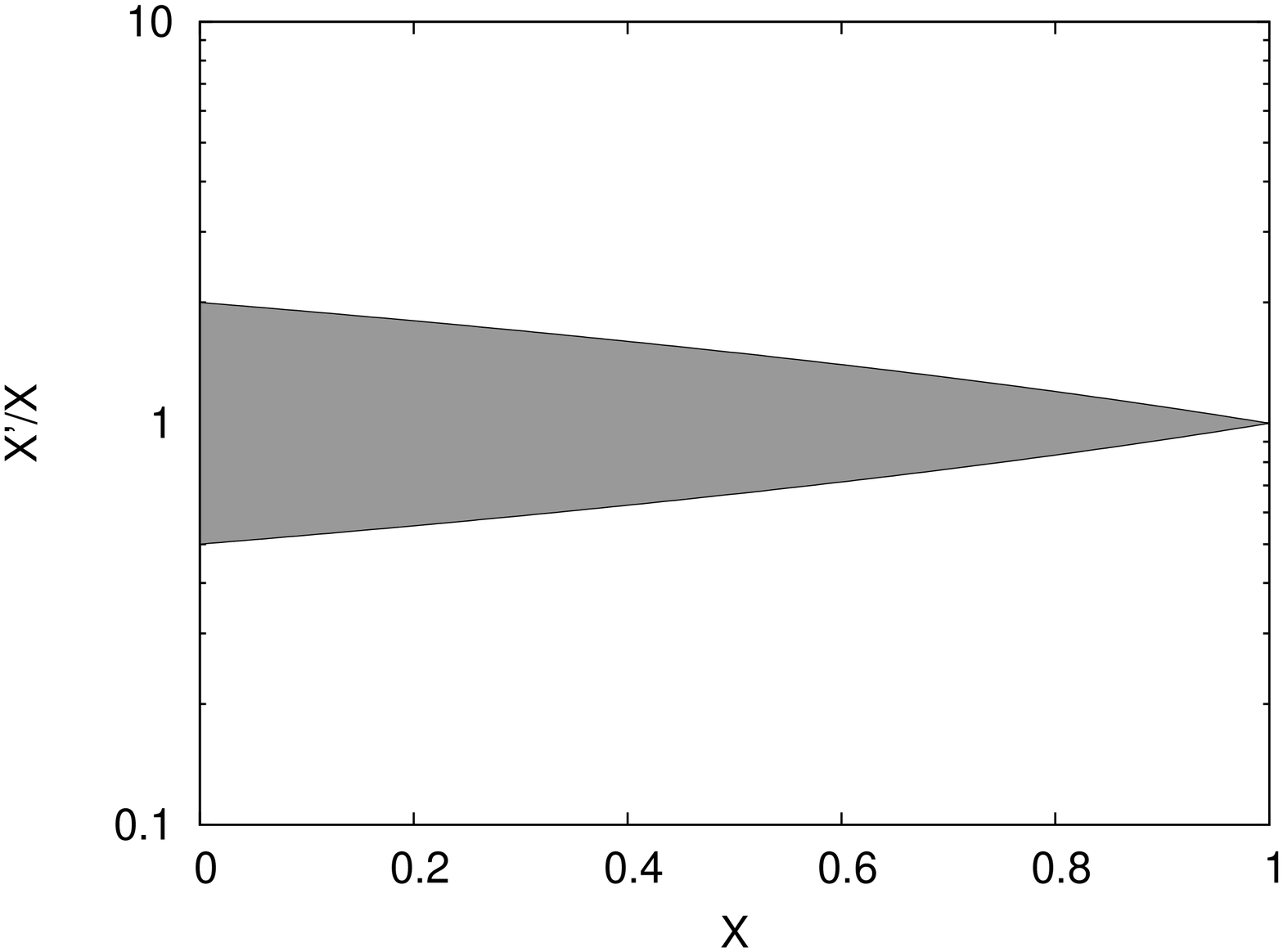}
\caption{\label{fig:xuncert}
Scaling of the uncertainty in $X$ with $X$. Shown is the ratio $X'/X$ when assuming an averaged uncertainty factor $\overline{u}$ of two. See text for a discussion.}
\end{figure}

With the above assumption we obtain
\begin{eqnarray}
u_{1/X}&=&1+\overline{u}\left\{ \frac{w_1}{w_0}\frac{R_1}{R_0} +\frac{w_2}{w_0}\frac{R_2}{R_0} +\frac{w_3}{w_0}\frac{R_3}{R_0} +\dots \right\} \nonumber \\
&=&1+\overline{u}\left\{ \frac{1}{X}-1\right\} \quad.
\end{eqnarray}
Then the factor $X'/X$ by which $X$ changes when assuming an averaged uncertainty factor $\overline{u}$ in the rate ratios is
\begin{equation}
u_X=\frac{X'}{X}=\frac{1}{\overline{u}\left(1-X\right) + X} \quad.
\end{equation}
This is an important result as it shows that the uncertainties will be negligible for $X\approx1$. Uncertainties will be larger for $X\ll1$
but will still yield $X\ll1$ when including them. In other words, large $X$ always remain large and very small $X$ remain small. Therefore
$X$ is a robust measure of the g.s.\ contribution and can be used to determine the direct impact of measurements of g.s.\ cross sections
on the astrophysical reaction rate. This is another advantage of using $X$ instead of the SEF. Figure \ref{fig:xuncert} shows the connection between the uncertainty and $X$.

The uncertainty factor will reach its maximum value $\max\left( \overline{u},1/\overline{u}\right)$ only for very small values of $X$.
It has to be kept in mind that $\overline{u}$ gives the uncertainty in the \textit{ratio} $R_i/R_0$ and not in the rates or cross sections. Global
reaction rate predictions typically find averaged deviations from measurements of 30\% for \rng\ MACS at 30 keV, with local deviations of up to factors $2-3$ \citep{rtk97,hoff99,Bao00}. When using locally adjusted parameters, these uncertainties can be reduced further (at the expense of predictive power for unknown nuclides and their properties). It is commonly assumed, however, that the cross section ratios $\sigma_i/\sigma_0$ and thus the reactivity ratios $R_i/R_0$ are predicted with much smaller uncertainty. This is especially true for the $s$ process. Unknown spins and parities of low-lying
states would give rise to the largest part of the uncertainty in the prediction of $\sigma_i/\sigma_0$ but these are known for nuclei at $s$ process conditions. The remaining uncertainty is dominated by the uncertainty in the energy dependence of the $\gamma$ widths \citep{energywindows}. However, it has to be realized that neutron captures in the $s$ process generally have large reaction $Q$ values and therefore the relative change in energy for the
contributing transitions $(E-E_i+Q)/Q$ \citep{raugamma,tommyreview} when varying $E_i$ is small. This leads to an uncertainty much smaller than the one in the specific cross sections themselves, probably better than 10\%.

\begin{figure}
\includegraphics[angle=-90,width=1\columnwidth]{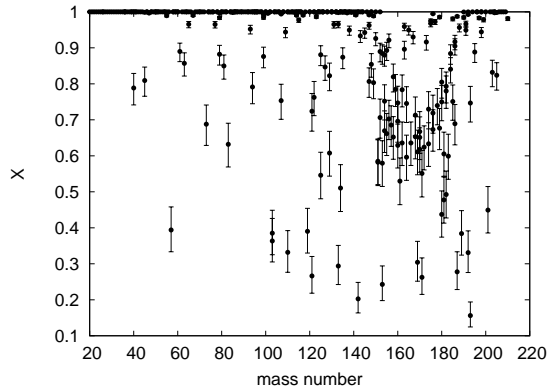}
\caption{
\label{fig:30comparison}
Ground state contributions $X$ to the stellar (n,$\gamma$) rate at $kT =
30$\,keV or $T = 0.384$\,GK for our set of nuclides for $s$ process calculations (see text). The errors are computed as described in Section \ref{sec:uncert}.}
\end{figure}

For a more conservative estimate, the uncertainties in $X$ in Table \ref{tab:results} and Figures \ref{fig:30comparison}, \ref{fig:2p5comparison} are given for uncertainty factors $\overline{u}=1.3$, i.e., 30\%.

\section{Reduction of model uncertainties in stellar rates}

\subsection{Maximally possible reduction by determining $\sigma_0$}

The g.s.\ contribution $X$ to the stellar rate, as derived in Section \ref{sec:gscontrib}, is not just of theoretical interest. It directly gives the maximum reduction in the rate uncertainty which can be possibly achieved by completely determining the g.s.\ rate factor, e.g., by completely measuring the g.s.\ cross sections $\sigma_0$ within the relevant energy window. Let us define an uncertainty factor $\mathcal{U}$ of the stellar rate $R^*$, implying that the "true" rate $R^*_\mathrm{true}$ is in the range $R^*/\mathcal{U}\leq R^*_\mathrm{true}\leq \mathcal{U}R^*$ for $\mathcal{U}\geq 1$. For instance, a 50\% uncertainty in the rate is then expressed by $\mathcal{U}=1.5$ and a rate known without uncertainty is characterized by $\mathcal{U}=1$. This overall uncertainty factor $\mathcal{U}$ of the rate can be split into a contribution to the uncertainty stemming from the g.s.\ uncertainty $\mathcal{U}_\mathrm{gs}$ and a combined uncertainty of the usually only theoretically predicted excited state contributions $\mathcal{U}_\mathrm{exc}$, leading to
\begin{equation}
\label{eq:maxreduct}
\mathcal{U}=\mathcal{U}_\mathrm{gs}X+\mathcal{U}_\mathrm{exc}\left( 1-X \right) \quad.
\end{equation}
(It is important to note that the uncertainty factors $\mathcal{U}$, $\mathcal{U}_\mathrm{gs}$, and $\mathcal{U}_\mathrm{exc}$ defined here are different from, but related to, the factors $\overline{u}=\mathcal{U}_\mathrm{exc}/\mathcal{U}_\mathrm{gs}$, $u_1$, $u_2$, \dots used in Section \ref{sec:uncert}.)

The uncertainty $\mathcal{U}$ is determined by $\mathcal{U}_\mathrm{gs}$ for $X\approx 1$ and by $\mathcal{U}_\mathrm{exc}$ for $X\ll 1$. In both cases it is safe to assume that the uncertainty of a theoretically predicted rate is $\mathcal{U}=\mathcal{U}_\mathrm{gs}=\mathcal{U}_\mathrm{exc}$ for simplicity.
If the g.s.\ cross section $\sigma_0$ can be experimentally determined without error, i.e.\ $\mathcal{U}_\mathrm{gs}^\mathrm{exp}=1$, within the full energy range required to compute the reaction rate integral, then the maximally possible percentual reduction of $\mathcal{U}$ will be simply $100 X$. It is immediately seen that the reduction is negligible for $X\ll 1$. On the other hand, if the experimentally determined g.s.\ rate $R_0$ has a non-negligible uncertainty $\mathcal{U}_\mathrm{gs}^\mathrm{exp}>1$, then the new, reduced uncertainty factor can be computed by replacing $\mathcal{U}_\mathrm{gs}$ in Equation (\ref{eq:maxreduct}) with $\mathcal{U}_\mathrm{gs}^\mathrm{exp}$. Again, this will only have appreciable impact on $\mathcal{U}$ when $X\approx 1$.

This maximally possible reduction in rate uncertainty is especially important for the $s$ process. Both the astrophysical modeling and the cross section measurements have entered a high-precision era, with $\sigma_0$ for many targets determined to a few percent precision and stellar models demanding such small uncertainties in the \textit{stellar} rates for a detailed understanding of the production of $s$ nuclei and for constraining the conditions within a star. It is important to realize, however, that the stellar rate is constrained with the experimental uncertainty only when $X=1$. In the extreme case $X \ll 1$ a measurement of g.s.\ cross sections or MACS will not be able to constrain the rate in any way, even when performed with the highest precision.
In this case $R^\ast$ is essentially given by the theoretical
prediction of the cross sections of the excited states, and further experiments
are required to constrain the theoretical predictions. For example, (n,n')
reactions for the population of the excited target states \citep[see, e.g.,][]{Mos10b} or \rgn\ reactions on the
residual nucleus \citep[e.g.,][]{sonn03branch} have been suggested to study specific transitions to excited states in the target.

It should be noted that even
with $X<1$ the knowledge of $\sigma_0$ can be used to test and improve the
model prediction of the reaction cross section, even though the measurement will not directly constrain $R^*$.  In many cases, deficiencies
in the description of the cross sections of g.s.\ and excited states are
correlated and the prediction of the stellar rate may be improved also in such
instances by a renormalization of the rate. Whether this is possible depends on reaction details and has to be
investigated separately for each target nucleus.

For completeness it has to be mentioned here that the uncertainties in the
SEFs $f_{\mathrm{SEF}}$ and ground state contributions $X$ are not the only
nuclear uncertainties playing a role in $s$-process nucleosynthesis
calculations. In particular, for the interesting branching nuclei often only theoretical
predictions of the capture cross sections are available. The feasibility of a considerable reduction of the theoretical uncertainties by
measurements can be checked by inspecting $X$ for such nuclei. But the
branching may be additionally affected by a temperature-dependent $\beta$-decay half-life
\citep{Tak87}. Also for decay rates (as for any other type of rate) the g.s.\ contribution can be determined by applying Equation (\ref{eq:X}).
In this work, however, we focus on neutron captures close to stability.

\subsection{General results}

\begin{figure}
\includegraphics[angle=-90,width=1\columnwidth]{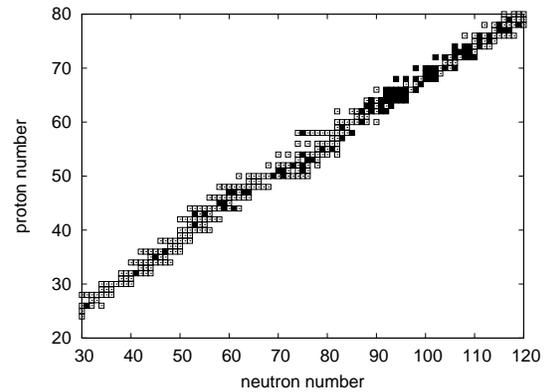}
\caption{\label{fig:surprise}
The open squares are a subset of the nuclides contained in our set for the $s$ process, having $X_\mathrm{lower} \geq 0.8$ for neutron capture (see Table \ref{tab:results} for the full list). The black squares mark all nuclei in our set for which $X_\mathrm{lower} < 0.8$ at $kT=30$ keV and therefore
an experimental measurement of g.s.\ cross sections cannot reduce the theoretical uncertainty by more than 80\%.}
\end{figure}

We have calculated SEFs $f_\mathrm{SEF}$ and g.s.\ contributions $X$ for a set of nuclides important
for $s$-process studies with the reaction rate code SMARAGD, version 0.8.1s \citep{tommyreview}. This includes the excited states from the recent version of \cite{ENSDF}. Nuclides present in KADoNiS v0.3 \citep{KADONIS} were supplemented with several additional nuclei under consideration
for inclusion in future KADoNiS versions, yielding a set of 411 nuclei. Table \ref{tab:results} shows the considered nuclei with their
values of $G_0$, $X$, and $f_\mathrm{SEF}$ at $kT=30$ keV and $kT=80$ keV (results for further temperatures will be included in future versions of KADoNiS). Also given is the relative uncertainty factor $u_X$ which defines the lower limit $X_\mathrm{lower}=X/u_X$ and upper limit $X_\mathrm{upper}=u_X X$ of $X$ as discussed in Section \ref{sec:uncert}. The last column of Table \ref{tab:results} provides information on the current experimental status of the \rng\ reaction on the given target nucleus and identifies the nuclei marked in Figure \ref{fig:surprise} (marked by an asterisk in the comment column). When a reaction is measured (marked by "e") and $X$ is close to unity, the uncertainty in the stellar rate is directly connected to the experimental uncertainty. If $X$ is small, the uncertainty stemming from theory has not been reduced even when the rate was measured. For several reactions, only a 30 keV MACS was measured (marked by "30") and the rates at other temperatures are determined from a renormalized rate prediction. Also listed are purely theoretical rates (marked by "t") and rates not appearing in KADoNiS v0.3 (marked by "n").

The table is useful for both astrophysicists and experimentalists and is meant to be interpreted as follows. The maximally possible reduction of the theory uncertainty can be obtained by inserting the value of $X$ into Equation (\ref{eq:maxreduct}) for a given reaction. Looking for reactions to better constrain experimentally, those with $X$ close to unity should be chosen. If they were measured previously, it would be possible to better constrain them by measuring with higher precision across the full relevant energy range \citep[see][]{energywindows}. An unmeasured rate will also only be constrained by measuring $\sigma_0$ if $X\approx1$. On the other extreme, some measured reactions show low values for $X$ (marked by the asterisk). It would be incorrect to assume for these cases that the uncertainty of the stellar rate is similar to the one of the measurement. Thus, this allows to assess which
reactions presently are well constrained in $s$ process calculations.

For a number of reactions, only $kT=30$ keV MACS are directly obtained from experiment and the energy dependence taken from theory. Obviously, this will only constrain the stellar rate at 30 keV when $X\approx1$. Although it is usually preferable to obtain a rate from a measurement within the full range of contributing energies, one has to realize that $X$ may decrease with higher energy, even when it is close to unity at $kT=30$ keV. In such a case, additional measurements above 30 keV may only partially reduce the uncertainty in the rates at higher temperature. In order to see the change in $X$ with energy, also values for $kT=80$ keV are shown in Table \ref{tab:results}.

\begin{figure}
\includegraphics[angle=-90,width=1\columnwidth]{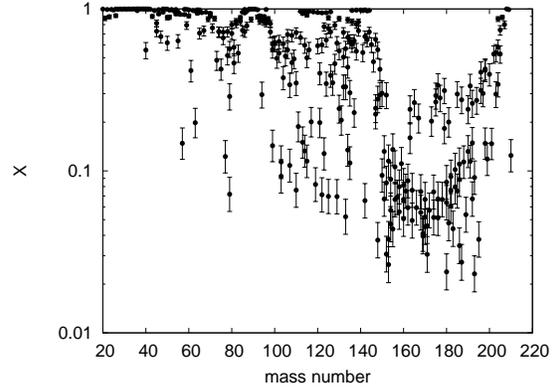}
\caption{
\label{fig:2p5comparison}
Same as figure \protect\ref{fig:30comparison} but for $T=2.5$ GK. Note the logarithmic scale.}
\end{figure}

For a better overview, further comparisons of g.s.\ contributions
$X$ to (n,$\gamma$) rates at different temperatures are shown in
Figures \ref{fig:30comparison} and \ref{fig:2p5comparison} for the nuclei included in Table \ref{tab:results}. The error bars on the $X$ values were
computed as described in Section \ref{sec:uncert}. We have chosen the reference temperature
$kT = 30$ keV for the $s$ process and a high temperature of $T = 2.5$ GK
typical for explosive nucleosynthesis.
As can be seen, the values of $X$ are smallest in the region of deformed
nuclei where also their deviation from the SEF is the largest. This is
explained by the fact that the level density in such nuclei is high, i.e.,
there are already many excited states at low excitation energy. Already at $kT
= 30$ keV we find $X \approx 0.16 - 0.3$ for several nuclei (with the lowest
values of $X=0.04$, $X=0.16$, and $X=0.2$ for $^{166}$Ho,
$^{193}$Pt, and $^{142}$Pr, respectively) and $X \approx 0.5
- 0.8$ for most nuclei in the mass range $150 \le A \le 190$. Although
experimental data with small uncertainties exist for most stable nuclei in
this mass range \citep{Bao00,KADONISV0,KADONIS}, the stellar reactivity
$R^\ast$ is at least partially based on the calculated contribution of excited
states. Consequently, the uncertainties of the stellar reactivities $R^\ast$
may be much larger than the uncertainties of the ground state MACS which
are provided in the compilations \citet{Bao00,KADONISV0,KADONIS}.

At a glance, Figure \ref{fig:surprise} provides a guide to which targets are well suited for experimentally constraining the
theory uncertainties in rates and which are not. The filled squares mark nuclei with $X<0.8$ (using the lower limit of $X$ from the errors
discussed in Section \ref{sec:uncert}). For neutron captures on these targets, even a complete experimental determination of the
g.s.\ rates will not be able to reduce the error from theory by more than 80\% and for even smaller $X$ it will not provide an improved
rate. The open squares mark nuclei in our set having $X\geq 0.8$ and therefore show cases for which the rate uncertainty mainly stems
from experimental uncertainties. If not measured already, they may provide good opportunities for future experiments.

The cut at 80\%
error reduction was chosen because a significant reduction in the uncertainty from theory is required in order to achieve the precision desired for
$s$-process nucleosynthesis. For example, assuming a 80\% reduction in an uncertainty of 30\% in the rate predictions \citep{rtk97},
the remaining error stemming from theory will then be of the order of 6\%. Assuming larger factors of $2-3$ in global
reaction rate predictions reduced by 80\% would lead to a theory uncertainty in the stellar rate of about $40-60$\% from the
excited state contributions. The experimental uncertainty for $\sigma_0$ has to be additionally included. This leads to uncertainties in stellar
rates which already may be too large for high-precision $s$-process nucleosynthesis models.

Neutron captures and their inverse photodisintegration reactions at higher temperatures are also important in the $\gamma$-process production of $p$ nuclei \citep{woohow,raudeflect}. For that reason, they are also included in KADoNiS \citep{KADONIS}.
It is apparent in Figure \ref{fig:2p5comparison} that $X$ can
assume very low values for higher temperature. This indicates that one has
to be very careful in extrapolating experimental rates from $s$-process
temperatures ($T=0.384$ GK is $E=30$ keV), e.g., to a typical $\gamma$-process
temperature of 2.5 GK. This implies an
increasing theory contribution which is only constrained by experiment, as
pointed out above, if the predictions of the g.s.\ and excited state cross
sections are correlated. Furthermore, while the partition function $G_0$ is
well determined at $s$-process conditions by experimentally known level
schemes, going to higher temperature (and/or unstable nuclei) implies an
increasing uncertainty also in $G_0$.

\subsection{Specific examples}
Finally, let us discuss some specific cases in more detail to further illustrate the difference between $f_\mathrm{SEF}$ and $X$: neutron
capture reactions on $^{79}$Se, $^{95}$Zr, $^{121}$Sn, $^{187}$Os, and
$^{193}$Pt. We focus here on the stellar temperatures $kT = 8$ keV and $kT =
30$ keV. A
temperature of $kT = 30$ keV is often taken as the reference temperature of
the $s$ process \citep[see, e.g.,][]{KADONIS}. It is close to the typical
burning temperature of the $^{22}$Ne\ran $^{25}$Mg reaction which is one of
the two neutron sources for the $s$ process in thermally pulsing AGB
stars. The main neutron source in AGB stars is $^{13}$C\ran $^{16}$O which
operates at lower temperatures around $kT = 8$ keV. Already at these
relatively low temperatures the g.s.\ contribution $X$ to the stellar reaction
rate factor $R^\ast$ is far below unity in some cases. The weak $s$ process in
massive stars operates at higher temperatures, and thus the ground state
contribution $X$ is even further reduced.

The reaction $^{79}$Se(n,$\gamma$)$^{80}$Se is an example of a case where
$f_\mathrm{SEF}<1$
due to a weak influence of exited states. The nucleus $^{79}$Se has several
low-lying levels ($J^\pi = 1/2^-$, 95.8\,keV;
$(1/2^-)$, 128\,keV(?); $9/2^+$, 137.0\,keV) above its $7/2^+$
g.s.~\citep{Singh02,ENSDF}.
Because of $J_0 = 7/2$ and $J_1 = 1/2$ of the first excited
state, its population remains below about
1\,\% up to $kT = 30$\,keV, and a similar value is found for the large $J = 9/2$
state at 137.0\,keV. This leads to $G_0=1.027$, $f_\mathrm{SEF}=0.99$, and
$X=0.98$ at $kT=30$ keV.
Moreover, $^{79}$Se is considered as an important
branching nucleus
with a highly temperature-dependent half-life \citep{Tak87}
in the weak $s$ process. For $kT=100$ keV we find
$f_\mathrm{SEF}=0.85$ and $X=0.78$. This means that
not only the excited states are populated more ($G_0=1.51$) but also the
captures on these states impact
the resulting stellar rate. Were this not the case, i.e., were the capture
cross sections of $^{79}$Se in the excited states negligible,
then $X=1$ and $f_\mathrm{SEF}=1/G_0=0.66$.

The reaction $^{95}$Zr(n,$\gamma$)$^{96}$Zr is an example of both
$f_\mathrm{SEF}=1$ and $X=1$ because excited states are not populated.
The first excited state of $^{95}$Zr is located at $E = 954$ keV
with $J^\pi = 1/2^+$. The g.s.\ has $J^\pi = 5/2^+$ \citep{ENSDF}. The SEF and
$X$ do not deviate from unity over the whole
temperature range of the
$s$ process because $G_0(T)=1$. Note that there was an earlier claim for a
low-lying first excited state with $J^\pi = (3/2,5/2)^+$ at 23\,keV
which has been observed only in one particular experiment \citep{Fro86}. The
existence of such a level has been excluded in a later high-resolution
experiment \citep{Sonn03}. There is no evidence for another low-lying state
``with quite different angular momentum'' as discussed recently in
\citet{Huth10}.  Consequently, the state at 23 keV has been removed from the
adopted levels of $^{95}$Zr \citep{Basu10,ENSDF}.  An experimental
determination of the $^{95}$Zr\rng $^{96}$Zr capture cross section would be
extremely valuable because the experimental neutron capture data completely
define the stellar reaction rate. These data are a prerequisite for the
understanding of isotopic patterns of zirconium and molybdenum isotopes in
meteorites which may be of $s$-process origin.

The reaction $^{121}$Sn\rng $^{122}$Sn is close to our example B above. The
g.s.\ of $^{121}$Sn has a low spin of
$J^\pi = 3/2^+$. There is a very low-lying intruder state at $E = 6.3$ keV
with $J^\pi = 11/2^-$, and a further low-$J$ state at $E = 60.3$ keV with
$J^\pi = 1/2^+$. Then there is a gap in the level scheme up to the next state
at $E = 663.6$ keV which is too high to play a significant role for the
$s$ process \citep{Ohya10,ENSDF}.
Already at $kT = 8$ keV the population of the first excited state exceeds
the g.s., and we find a g.s.\ contribution to the stellar rate
of $X = 0.36$. % (1.19 * 2.364)^-1
At the reference temperature of $kT = 30$ keV it is further reduced to $X =
0.27$. % (1.07 * 3.498)^-1
The SEFs remain close to unity with $f_\mathrm{SEF} = 1.19$ and 1.07 at
8\,keV and 30\,keV.
Because of the dominating contribution from the first excited state,
experimental data for the $^{121}$Sn\rng $^{122}$Sn reaction can only
determine a minor contribution of one third to the stellar rate.

The reaction $^{187}$Os\rng $^{188}$Os is important for the Re/Os
cosmochronometer \citep{Mos10a,Mos10b}.
The nucleus $^{187}$Os exhibits a very low-lying $3/2^-$
state at $9.8$ keV above the $1/2^-$ g.s. The next levels appear at
74.4 keV, 75.0 keV, and 100.5 keV ($3/2^-$, $5/2^-$, $7/2^-$). Above a level at
117 keV without spin assignment the next levels are found at 187.4 keV
and 190.6 keV, which is too high for the $s$ process \citep{Bas10,ENSDF}. At
$kT = 8$ keV the SEF is $f_\mathrm{SEF} = 1.0$. The first excited state,
however, is already populated and we find a g.s.\ contribution of $X
= 0.63$. % ( 1.591 )^-1
At higher energies the g.s.\ contribution reduces further to $X = 0.28$
% ( 1.19 * 3.014 )^-1
for $kT = 30$ keV, although $f_\mathrm{SEF} = 1.19$ remains close to unity.
Because of the relevance of this reaction for the
Re/Os cosmochronometer, neutron capture experiments on $^{187}$Os have been
performed \citep[see, e.g.,][and references therein]{Mos10a,Fuj10}. The
importance of the low-lying first excited state has been noticed in that work and
additional experiments on inelastic neutron scattering of $^{187}$Os and
photodisintegration of $^{188}$Os have been performed
\citep[see, e.g.,][]{Shi05,Mos10b} to study transitions proceeding on this state.

Another example for the importance of including excited state transitions is
$^{193}$Pt\rng $^{194}$Pt.
An extremely low-lying $3/2^-$ state at $E =
1.6$ keV is found above the $1/2^-$ g.s.\ of $^{193}$Pt, and a further
low-lying $5/2^-$ state is located at $E = 14.3$ keV. Then there is a small
gap in the level scheme up to the next $3/2^-$ state at $E = 114.2$ keV
\citep{Ach06,ENSDF}. Already at $kT = 8$ keV the first excited state is
dominating the reaction rate factor $R^\ast$ and the g.s.\
contribution is only $X = 0.26$. % ( 1.22 * 3.133 )^-1
At $kT = 30$ keV the g.s.\ contribution drops to $X = 0.16$,
% ( 1.31 * 4.880 )^-1
and the reactivity $R^\ast$ is essentially determined by comparable
contributions of the states at 1.6 keV and 14.3 keV. Again, as in the
previous examples, the SEF remains relatively close to unity with
$f_\mathrm{SEF} = 1.22$ at $kT = 8$ keV and $f_\mathrm{SEF} = 1.31$ at $kT =
30$ keV. Thus,
an experimental determination of the $^{193}$Pt\rng $^{194}$Pt cross
section can only be used to restrict theoretical predictions for the
g.s.\ cross section.
Similarly to the above examples $^{121}$Sn and
$^{187}$Os, it is not possible to directly derive the stellar reaction rate
from experimental neutron capture data.

\section{Summary}
In conclusion, SEFs close to unity should not be interpreted as that thermally
excited states play only a minor role in the stellar reaction rate. The
g.s.\ contribution $X$ to the stellar reaction rate -- given by $X =
(f_\mathrm{SEF} \, G_0)^{-1}$ -- may be significantly below 0.5 for target nuclei
with low-lying excited states also for cases with $f_\mathrm{SEF} \approx
1$. Low values of $X$ indicate that it is impossible to determine the stellar
reaction rate directly from experimental neutron capture data. This should be
taken into account in the planning of future neutron capture experiments, in
particular for the extremely difficult experiments on unstable branching
nuclei in the $s$ process. Because of the importance of the g.s.\ contribution
$X$, it will be included in future versions of the KADoNiS database
\citep{KADONISV0,KADONIS}.

Although $X$ is a theoretical quantity, it was shown that the
uncertainties of $X$ remain relatively small for most nuclei in the
$s$-process path because the partition functions can be calculated from
the experimentally well-known level scheme for these nuclei.

We showed that $X$ is a more useful quantity than the SEF. Experimentalists may still use both $X$ and SEF: the value of $X$ shows whether an experiment measuring g.s.\ cross sections or rates actually contributes to the calculation of the stellar rate; the SEF can be used to renormalize the g.s.\ rate to obtain a stellar rate. Such a renormalization, however, only makes sense when $X$ is close to unity. For modelers performing stellar network calculations, only $X$ provides valuable information. Usually, modelers assume that experimentally known rates are better constrained than theoretical ones or, at least, constrained within the experimental error. However, this would only be the case for $X=1$. A low value of $X$ for a reaction indicates that the modeler may allow for a larger uncertainty (for example, in a rate variation study), using Equation (\ref{eq:maxreduct}) with appropriate experimental and theoretical uncertainties.

Obviously, these findings apply not only to neutron capture but to any kind of
reaction, also in other nucleosynthesis processes. Especially at the high
temperatures of explosive nucleosynthesis, reactions on nuclei in excited
states often contribute significantly to the stellar rate. Only the
g.s.\ contribution $X$, which can be calculated in the same manner for all types of reactions,
allows to judge whether measurements of reactions on
nuclei in the g.s.\ will be able to put direct constraints on the stellar rate.

\acknowledgments
This work was supported by OTKA (NN83261). I.D.\ and R.P.\ are
supported by the Helmholtz association in the Young Investigators projects
VH-NG-627 and VH-NG-327. T.R. receives support from the THEXO workpackage of the ENSAR project within the European Commission FP7.

\clearpage

\clearpage

\begin{deluxetable}{rrlrrrrrrrrc}
\tablecaption{\label{tab:results}Normalized partition functions $G_0$, ground state contributions $X$, their uncertainty factors $u_X$, and stellar enhancement factors $f_\mathrm{SEF}$
at $kT=30$ keV  and $kT=80$ keV for nuclides identified by their charge number $Z$ and mass number $A$.}
%\tablewidth{0pt}
\tablehead{
\colhead{$Z$} & \colhead{$A$} & \colhead{Nucl.} & \colhead{$G_0^{30}$} & \colhead{$X^{30}$} & \colhead{$u_X^{30}$} &
\colhead{$f_\mathrm{SEF}^{30}$} & \colhead{$G_0^{80}$} & \colhead{$X^{80}$} & \colhead{$u_X^{80}$} &
\colhead{$f_\mathrm{SEF}^{80}$} & \colhead{Comment\tablenotemark{a}}
}
\startdata
 10 &  20 & $^{ 20}$Ne & 1.000 & 1.000 & 1.000 & 1.000 & 1.000 & 1.000 & 1.000 & 1.000  & e   \\
 10 &  21 & $^{ 21}$Ne & 1.000 & 1.000 & 1.000 & 1.000 & 1.019 & 0.993 & 1.002 & 0.988  & e   \\
 10 &  22 & $^{ 22}$Ne & 1.000 & 1.000 & 1.000 & 1.000 & 1.000 & 1.000 & 1.000 & 1.000  & 30   \\
 11 &  23 & $^{ 23}$Na & 1.000 & 1.000 & 1.000 & 1.000 & 1.006 & 0.997 & 1.001 & 0.997  & e   \\
 12 &  24 & $^{ 24}$Mg & 1.000 & 1.000 & 1.000 & 1.000 & 1.000 & 1.000 & 1.000 & 1.000  & e   \\
 12 &  25 & $^{ 25}$Mg & 1.000 & 1.000 & 1.000 & 1.000 & 1.000 & 1.000 & 1.000 & 1.000  & e   \\
 12 &  26 & $^{ 26}$Mg & 1.000 & 1.000 & 1.000 & 1.000 & 1.000 & 1.000 & 1.000 & 1.000  & e   \\
 13 &  26 & $^{ 26}$Al & 1.000 & 1.000 & 1.000 & 1.000 & 1.009 & 0.995 & 1.001 & 0.996  & t   \\
 13 &  27 & $^{ 27}$Al & 1.000 & 1.000 & 1.000 & 1.000 & 1.000 & 1.000 & 1.000 & 1.000  & e   \\
 14 &  28 & $^{ 28}$Si & 1.000 & 1.000 & 1.000 & 1.000 & 1.000 & 1.000 & 1.000 & 1.000  & e   \\
 14 &  29 & $^{ 29}$Si & 1.000 & 1.000 & 1.000 & 1.000 & 1.000 & 1.000 & 1.000 & 1.000  & e   \\
 14 &  30 & $^{ 30}$Si & 1.000 & 1.000 & 1.000 & 1.000 & 1.000 & 1.000 & 1.000 & 1.000  & e   \\
 15 &  31 & $^{ 31}$P  & 1.000 & 1.000 & 1.000 & 1.000 & 1.000 & 1.000 & 1.000 & 1.000  & e   \\
 15 &  32 & $^{ 32}$P  & 1.123 & 0.945 & 1.013 & 0.943 & 1.628 & 0.719 & 1.069 & 0.855  & n   \\
 15 &  33 & $^{ 33}$P  & 1.000 & 1.000 & 1.000 & 1.000 & 1.000 & 1.000 & 1.000 & 1.000  & n   \\
 16 &  32 & $^{ 32}$S  & 1.000 & 1.000 & 1.000 & 1.000 & 1.000 & 1.000 & 1.000 & 1.000  & e   \\
 16 &  33 & $^{ 33}$S  & 1.000 & 1.000 & 1.000 & 1.000 & 1.000 & 1.000 & 1.000 & 1.000  & e   \\
 16 &  34 & $^{ 34}$S  & 1.000 & 1.000 & 1.000 & 1.000 & 1.000 & 1.000 & 1.000 & 1.000  & e   \\
 16 &  35 & $^{ 35}$S  & 1.000 & 1.000 & 1.000 & 1.000 & 1.000 & 1.000 & 1.000 & 1.000  & n   \\
 16 &  36 & $^{ 36}$S  & 1.000 & 1.000 & 1.000 & 1.000 & 1.000 & 1.000 & 1.000 & 1.000  & e   \\
 17 &  35 & $^{ 35}$Cl & 1.000 & 1.000 & 1.000 & 1.000 & 1.000 & 1.000 & 1.000 & 1.000  & e   \\
 17 &  36 & $^{ 36}$Cl & 1.000 & 1.000 & 1.000 & 1.000 & 1.000 & 1.000 & 1.000 & 1.000  & t   \\
 17 &  37 & $^{ 37}$Cl & 1.000 & 1.000 & 1.000 & 1.000 & 1.000 & 1.000 & 1.000 & 1.000  & e   \\
 18 &  36 & $^{ 36}$Ar & 1.000 & 1.000 & 1.000 & 1.000 & 1.000 & 1.000 & 1.000 & 1.000  & t   \\
 18 &  37 & $^{ 37}$Ar & 1.000 & 1.000 & 1.000 & 1.000 & 1.000 & 1.000 & 1.000 & 1.000  & n   \\
 18 &  38 & $^{ 38}$Ar & 1.000 & 1.000 & 1.000 & 1.000 & 1.000 & 1.000 & 1.000 & 1.000  & t   \\
 18 &  39 & $^{ 39}$Ar & 1.000 & 1.000 & 1.000 & 1.000 & 1.000 & 1.000 & 1.000 & 1.000  & t   \\
 18 &  40 & $^{ 40}$Ar & 1.000 & 1.000 & 1.000 & 1.000 & 1.000 & 1.000 & 1.000 & 1.000  & 30  \\
 19 &  39 & $^{ 39}$K  & 1.000 & 1.000 & 1.000 & 1.000 & 1.000 & 1.000 & 1.000 & 1.000  & e   \\
 19 &  40 & $^{ 40}$K  & 1.288 & 0.788 & 1.051 & 0.985 & 1.536 & 0.636 & 1.092 & 1.024  & t,$\ast$   \\
 19 &  41 & $^{ 41}$K  & 1.000 & 1.000 & 1.000 & 1.000 & 1.000 & 1.000 & 1.000 & 1.000  & e   \\
 20 &  40 & $^{ 40}$Ca & 1.000 & 1.000 & 1.000 & 1.000 & 1.000 & 1.000 & 1.000 & 1.000  & 30   \\
 20 &  41 & $^{ 41}$Ca & 1.000 & 1.000 & 1.000 & 1.000 & 1.000 & 1.000 & 1.000 & 1.000  & t   \\
 20 &  42 & $^{ 42}$Ca & 1.000 & 1.000 & 1.000 & 1.000 & 1.000 & 1.000 & 1.000 & 1.000  & e   \\
 20 &  43 & $^{ 43}$Ca & 1.000 & 1.000 & 1.000 & 1.000 & 1.007 & 0.995 & 1.001 & 0.998  & e   \\
 20 &  44 & $^{ 44}$Ca & 1.000 & 1.000 & 1.000 & 1.000 & 1.000 & 1.000 & 1.000 & 1.000  & e   \\
 20 &  45 & $^{ 45}$Ca & 1.002 & 0.999 & 1.000 & 0.999 & 1.085 & 0.954 & 1.011 & 0.966  & t   \\
 20 &  46 & $^{ 46}$Ca & 1.000 & 1.000 & 1.000 & 1.000 & 1.000 & 1.000 & 1.000 & 1.000  & e   \\
 20 &  47 & $^{ 47}$Ca & 1.000 & 1.000 & 1.000 & 1.000 & 1.000 & 1.000 & 1.000 & 1.000  & n   \\
 20 &  48 & $^{ 48}$Ca & 1.000 & 1.000 & 1.000 & 1.000 & 1.000 & 1.000 & 1.000 & 1.000  & e   \\
 21 &  45 & $^{ 45}$Sc & 1.331 & 0.809 & 1.046 & 0.929 & 1.434 & 0.765 & 1.057 & 0.912  & e,$\ast$   \\
 21 &  46 & $^{ 46}$Sc & 1.259 & 0.917 & 1.019 & 0.866 & 1.910 & 0.737 & 1.065 & 0.711  & n   \\
 21 &  47 & $^{ 47}$Sc & 1.000 & 1.000 & 1.000 & 1.000 & 1.000 & 1.000 & 1.000 & 1.000  & n   \\
 21 &  48 & $^{ 48}$Sc & 1.011 & 0.988 & 1.003 & 1.001 & 1.199 & 0.798 & 1.049 & 1.045  & n   \\
 22 &  46 & $^{ 46}$Ti & 1.000 & 1.000 & 1.000 & 1.000 & 1.000 & 1.000 & 1.000 & 1.000  & e   \\
 22 &  47 & $^{ 47}$Ti & 1.007 & 0.997 & 1.001 & 0.997 & 1.187 & 0.906 & 1.022 & 0.930  & e   \\
 22 &  48 & $^{ 48}$Ti & 1.000 & 1.000 & 1.000 & 1.000 & 1.000 & 1.000 & 1.000 & 1.000  & e   \\
 22 &  49 & $^{ 49}$Ti & 1.000 & 1.000 & 1.000 & 1.000 & 1.000 & 1.000 & 1.000 & 1.000  & e   \\
 22 &  50 & $^{ 50}$Ti & 1.000 & 1.000 & 1.000 & 1.000 & 1.000 & 1.000 & 1.000 & 1.000  & e   \\
 23 &  50 & $^{ 50}$V  & 1.000 & 1.000 & 1.000 & 1.000 & 1.072 & 0.939 & 1.014 & 0.994  & t   \\
 23 &  51 & $^{ 51}$V  & 1.000 & 1.000 & 1.000 & 1.000 & 1.015 & 0.992 & 1.002 & 0.993  & e   \\
 24 &  50 & $^{ 50}$Cr & 1.000 & 1.000 & 1.000 & 1.000 & 1.000 & 1.000 & 1.000 & 1.000  & e   \\
 24 &  51 & $^{ 51}$Cr & 1.000 & 1.000 & 1.000 & 1.000 & 1.000 & 1.000 & 1.000 & 1.000  & t   \\
 24 &  52 & $^{ 52}$Cr & 1.000 & 1.000 & 1.000 & 1.000 & 1.000 & 1.000 & 1.000 & 1.000  & e   \\
 24 &  53 & $^{ 53}$Cr & 1.000 & 1.000 & 1.000 & 1.000 & 1.000 & 1.000 & 1.000 & 1.000  & e   \\
 24 &  54 & $^{ 54}$Cr & 1.000 & 1.000 & 1.000 & 1.000 & 1.000 & 1.000 & 1.000 & 1.000  & e   \\
 25 &  55 & $^{ 55}$Mn & 1.020 & 0.989 & 1.002 & 0.991 & 1.276 & 0.859 & 1.034 & 0.912  & e   \\
 26 &  54 & $^{ 54}$Fe & 1.000 & 1.000 & 1.000 & 1.000 & 1.000 & 1.000 & 1.000 & 1.000  & e   \\
 26 &  55 & $^{ 55}$Fe & 1.000 & 1.000 & 1.000 & 1.000 & 1.003 & 0.999 & 1.000 & 0.998  & t   \\
 26 &  56 & $^{ 56}$Fe & 1.000 & 1.000 & 1.000 & 1.000 & 1.000 & 1.000 & 1.000 & 1.000  & e   \\
 26 &  57 & $^{ 57}$Fe & 2.269 & 0.394 & 1.163 & 1.118 & 3.236 & 0.248 & 1.210 & 1.246  & e,$\ast$   \\
 26 &  58 & $^{ 58}$Fe & 1.000 & 1.000 & 1.000 & 1.000 & 1.000 & 1.000 & 1.000 & 1.000  & e   \\
 26 &  60 & $^{ 60}$Fe & 1.000 & 1.000 & 1.000 & 1.000 & 1.000 & 1.000 & 1.000 & 1.000  & 30  \\
 27 &  59 & $^{ 59}$Co & 1.000 & 1.000 & 1.000 & 1.000 & 1.000 & 1.000 & 1.000 & 1.000  & e   \\
 27 &  60 & $^{ 60}$Co & 1.065 & 0.909 & 1.022 & 1.034 & 1.268 & 0.746 & 1.062 & 1.058  & n   \\
 28 &  58 & $^{ 58}$Ni & 1.000 & 1.000 & 1.000 & 1.000 & 1.000 & 1.000 & 1.000 & 1.000  & e   \\
 28 &  59 & $^{ 59}$Ni & 1.000 & 1.000 & 1.000 & 1.000 & 1.023 & 0.986 & 1.003 & 0.991  & t   \\
 28 &  60 & $^{ 60}$Ni & 1.000 & 1.000 & 1.000 & 1.000 & 1.000 & 1.000 & 1.000 & 1.000  & e   \\
 28 &  61 & $^{ 61}$Ni & 1.159 & 0.890 & 1.026 & 0.970 & 1.661 & 0.624 & 1.095 & 0.966  & e   \\
 28 &  62 & $^{ 62}$Ni & 1.000 & 1.000 & 1.000 & 1.000 & 1.000 & 1.000 & 1.000 & 1.000  & 30  \\
 28 &  63 & $^{ 63}$Ni & 1.175 & 0.857 & 1.034 & 0.993 & 2.298 & 0.437 & 1.149 & 0.996  & t   \\
 28 &  64 & $^{ 64}$Ni & 1.000 & 1.000 & 1.000 & 1.000 & 1.000 & 1.000 & 1.000 & 1.000  & e   \\
 29 &  63 & $^{ 63}$Cu & 1.000 & 1.000 & 1.000 & 1.000 & 1.000 & 1.000 & 1.000 & 1.000  & e   \\
 29 &  64 & $^{ 64}$Cu & 1.008 & 0.992 & 1.002 & 0.999 & 1.321 & 0.790 & 1.051 & 0.958  & n   \\
 29 &  65 & $^{ 65}$Cu & 1.000 & 1.000 & 1.000 & 1.000 & 1.000 & 1.000 & 1.000 & 1.000  & e   \\
 30 &  64 & $^{ 64}$Zn & 1.000 & 1.000 & 1.000 & 1.000 & 1.000 & 1.000 & 1.000 & 1.000  & e   \\
 30 &  65 & $^{ 65}$Zn & 1.070 & 0.966 & 1.008 & 0.968 & 1.378 & 0.863 & 1.033 & 0.840  & t   \\
 30 &  66 & $^{ 66}$Zn & 1.000 & 1.000 & 1.000 & 1.000 & 1.000 & 1.000 & 1.000 & 1.000  & e   \\
 30 &  67 & $^{ 67}$Zn & 1.016 & 0.990 & 1.002 & 0.994 & 1.176 & 0.924 & 1.018 & 0.920  & e   \\
 30 &  68 & $^{ 68}$Zn & 1.000 & 1.000 & 1.000 & 1.000 & 1.000 & 1.000 & 1.000 & 1.000  & e   \\
 30 &  69 & $^{ 69}$Zn & 1.000 & 1.000 & 1.000 & 1.000 & 1.025 & 0.975 & 1.006 & 1.000  & n   \\
 30 &  70 & $^{ 70}$Zn & 1.000 & 1.000 & 1.000 & 1.000 & 1.000 & 1.000 & 1.000 & 1.000  & t   \\
 31 &  69 & $^{ 69}$Ga & 1.000 & 1.000 & 1.000 & 1.000 & 1.010 & 0.996 & 1.001 & 0.994  & e   \\
 31 &  70 & $^{ 70}$Ga & 1.000 & 1.000 & 1.000 & 1.000 & 1.004 & 0.997 & 1.001 & 0.999  & n   \\
 31 &  71 & $^{ 71}$Ga & 1.000 & 1.000 & 1.000 & 1.000 & 1.010 & 0.996 & 1.001 & 0.995  & 30   \\
 32 &  70 & $^{ 70}$Ge & 1.000 & 1.000 & 1.000 & 1.000 & 1.000 & 1.000 & 1.000 & 1.000  & e   \\
 32 &  71 & $^{ 71}$Ge & 1.016 & 0.982 & 1.004 & 1.002 & 1.767 & 0.530 & 1.122 & 1.067  & n   \\
 32 &  72 & $^{ 72}$Ge & 1.000 & 1.000 & 1.000 & 1.000 & 1.000 & 1.000 & 1.000 & 1.000  & t   \\
 32 &  73 & $^{ 73}$Ge & 1.488 & 0.688 & 1.078 & 0.977 & 1.952 & 0.575 & 1.109 & 0.891  & t,$\ast$   \\
 32 &  74 & $^{ 74}$Ge & 1.000 & 1.000 & 1.000 & 1.000 & 1.003 & 0.996 & 1.001 & 1.001  & 30   \\
 32 &  75 & $^{ 75}$Ge & 1.050 & 0.956 & 1.010 & 0.996 & 2.536 & 0.463 & 1.141 & 0.851  & n   \\
 32 &  76 & $^{ 76}$Ge & 1.000 & 1.000 & 1.000 & 1.000 & 1.004 & 0.996 & 1.001 & 1.000  & 30   \\
 33 &  75 & $^{ 75}$As & 1.001 & 1.000 & 1.000 & 0.999 & 1.193 & 0.900 & 1.024 & 0.931  & e   \\
 33 &  76 & $^{ 76}$As & 1.198 & 0.910 & 1.021 & 0.917 & 2.168 & 0.640 & 1.091 & 0.721  & n   \\
 33 &  77 & $^{ 77}$As & 1.002 & 0.999 & 1.000 & 0.998 & 1.219 & 0.920 & 1.019 & 0.891  & n   \\
 34 &  74 & $^{ 74}$Se & 1.000 & 1.000 & 1.000 & 1.000 & 1.002 & 0.997 & 1.001 & 1.001  & 30   \\
 34 &  75 & $^{ 75}$Se & 1.051 & 0.968 & 1.007 & 0.983 & 1.677 & 0.696 & 1.076 & 0.858  & n   \\
 34 &  76 & $^{ 76}$Se & 1.000 & 1.000 & 1.000 & 1.000 & 1.005 & 0.994 & 1.001 & 1.001  & e   \\
 34 &  77 & $^{ 77}$Se & 1.034 & 0.965 & 1.008 & 1.002 & 2.409 & 0.430 & 1.151 & 0.965  & t   \\
 34 &  78 & $^{ 78}$Se & 1.000 & 1.000 & 1.000 & 1.000 & 1.002 & 0.997 & 1.001 & 1.001  & 30   \\
 34 &  79 & $^{ 79}$Se & 1.027 & 0.984 & 1.004 & 0.990 & 1.361 & 0.836 & 1.039 & 0.879  & t   \\
 34 &  80 & $^{ 80}$Se & 1.000 & 1.000 & 1.000 & 1.000 & 1.001 & 0.999 & 1.000 & 1.000  & e   \\
 34 &  81 & $^{ 81}$Se & 1.129 & 0.848 & 1.036 & 1.044 & 2.244 & 0.356 & 1.175 & 1.253  & n   \\
 34 &  82 & $^{ 82}$Se & 1.000 & 1.000 & 1.000 & 1.000 & 1.001 & 0.999 & 1.000 & 1.000  & t   \\
 35 &  79 & $^{ 79}$Br & 1.004 & 0.997 & 1.001 & 1.000 & 1.354 & 0.763 & 1.058 & 0.968  & e   \\
 35 &  80 & $^{ 80}$Br & 1.696 & 0.582 & 1.107 & 1.012 & 3.893 & 0.226 & 1.217 & 1.136  & n,$\ast$   \\
 35 &  81 & $^{ 81}$Br & 1.000 & 1.000 & 1.000 & 1.000 & 1.053 & 0.961 & 1.009 & 0.989  & e   \\
 35 &  82 & $^{ 82}$Br & 1.121 & 0.904 & 1.023 & 0.987 & 1.397 & 0.777 & 1.054 & 0.921  & n   \\
 36 &  78 & $^{ 78}$Kr & 1.000 & 1.000 & 1.000 & 1.000 & 1.017 & 0.978 & 1.005 & 1.006  & e   \\
 36 &  79 & $^{ 79}$Kr & 1.115 & 0.882 & 1.028 & 1.017 & 3.383 & 0.265 & 1.204 & 1.115  & t   \\
 36 &  80 & $^{ 80}$Kr & 1.000 & 1.000 & 1.000 & 1.000 & 1.002 & 0.997 & 1.001 & 1.001  & e   \\
 36 &  81 & $^{ 81}$Kr & 1.240 & 0.850 & 1.036 & 0.949 & 1.700 & 0.637 & 1.092 & 0.924  & t   \\
 36 &  82 & $^{ 82}$Kr & 1.000 & 1.000 & 1.000 & 1.000 & 1.000 & 1.000 & 1.000 & 1.000  & e   \\
 36 &  83 & $^{ 83}$Kr & 1.635 & 0.632 & 1.093 & 0.968 & 1.831 & 0.564 & 1.112 & 0.968  & e,$\ast$   \\
 36 &  84 & $^{ 84}$Kr & 1.000 & 1.000 & 1.000 & 1.000 & 1.000 & 1.000 & 1.000 & 1.000  & e   \\
 36 &  85 & $^{ 85}$Kr & 1.000 & 1.000 & 1.000 & 1.000 & 1.004 & 0.996 & 1.001 & 1.000  & t   \\
 36 &  86 & $^{ 86}$Kr & 1.000 & 1.000 & 1.000 & 1.000 & 1.000 & 1.000 & 1.000 & 1.000  & e   \\
 37 &  85 & $^{ 85}$Rb & 1.004 & 0.998 & 1.000 & 0.997 & 1.113 & 0.956 & 1.010 & 0.939  & e   \\
 37 &  86 & $^{ 86}$Rb & 1.000 & 1.000 & 1.000 & 1.000 & 1.005 & 0.996 & 1.001 & 0.999  & t   \\
 37 &  87 & $^{ 87}$Rb & 1.000 & 1.000 & 1.000 & 1.000 & 1.010 & 0.993 & 1.002 & 0.997  & e   \\
 38 &  84 & $^{ 84}$Sr & 1.000 & 1.000 & 1.000 & 1.000 & 1.000 & 1.000 & 1.000 & 1.000  & 30   \\
 38 &  85 & $^{ 85}$Sr & 1.000 & 1.000 & 1.000 & 1.000 & 1.054 & 0.966 & 1.008 & 0.981  & n   \\
 38 &  86 & $^{ 86}$Sr & 1.000 & 1.000 & 1.000 & 1.000 & 1.000 & 1.000 & 1.000 & 1.000  & e   \\
 38 &  87 & $^{ 87}$Sr & 1.000 & 1.000 & 1.000 & 1.000 & 1.002 & 0.999 & 1.000 & 1.000  & e   \\
 38 &  88 & $^{ 88}$Sr & 1.000 & 1.000 & 1.000 & 1.000 & 1.000 & 1.000 & 1.000 & 1.000  & e   \\
 38 &  89 & $^{ 89}$Sr & 1.000 & 1.000 & 1.000 & 1.000 & 1.000 & 1.000 & 1.000 & 1.000  & t   \\
 38 &  90 & $^{ 90}$Sr & 1.000 & 1.000 & 1.000 & 1.000 & 1.000 & 1.000 & 1.000 & 1.000  & n   \\
 39 &  89 & $^{ 89}$Y  & 1.000 & 1.000 & 1.000 & 1.000 & 1.000 & 1.000 & 1.000 & 1.000  & e   \\
 39 &  90 & $^{ 90}$Y  & 1.002 & 0.999 & 1.000 & 1.000 & 1.112 & 0.918 & 1.019 & 0.980  & n   \\
 39 &  91 & $^{ 91}$Y  & 1.000 & 1.000 & 1.000 & 1.000 & 1.005 & 0.992 & 1.002 & 1.003  & n   \\
 40 &  90 & $^{ 90}$Zr & 1.000 & 1.000 & 1.000 & 1.000 & 1.000 & 1.000 & 1.000 & 1.000  & e   \\
 40 &  91 & $^{ 91}$Zr & 1.000 & 1.000 & 1.000 & 1.000 & 1.000 & 1.000 & 1.000 & 1.000  & 30   \\
 40 &  92 & $^{ 92}$Zr & 1.000 & 1.000 & 1.000 & 1.000 & 1.000 & 1.000 & 1.000 & 1.000  & e   \\
 40 &  93 & $^{ 93}$Zr & 1.000 & 1.000 & 1.000 & 1.000 & 1.024 & 0.992 & 1.002 & 0.985  & e   \\
 40 &  94 & $^{ 94}$Zr & 1.000 & 1.000 & 1.000 & 1.000 & 1.000 & 1.000 & 1.000 & 1.000  & e   \\
 40 &  95 & $^{ 95}$Zr & 1.000 & 1.000 & 1.000 & 1.000 & 1.001 & 0.999 & 1.000 & 1.000  & t   \\
 40 &  96 & $^{ 96}$Zr & 1.000 & 1.000 & 1.000 & 1.000 & 1.000 & 1.000 & 1.000 & 1.000  & e   \\
 41 &  93 & $^{ 93}$Nb & 1.072 & 0.952 & 1.011 & 0.980 & 1.136 & 0.915 & 1.020 & 0.962  & e   \\
 41 &  94 & $^{ 94}$Nb & 1.342 & 0.791 & 1.051 & 0.942 & 2.382 & 0.493 & 1.132 & 0.851  & t,$\ast$   \\
 41 &  95 & $^{ 95}$Nb & 1.000 & 1.000 & 1.000 & 1.000 & 1.011 & 0.990 & 1.002 & 0.999  & t   \\
 41 &  96 & $^{ 96}$Nb & 1.201 & 0.862 & 1.033 & 0.966 & 1.716 & 0.625 & 1.095 & 0.933  & n   \\
 42 &  92 & $^{ 92}$Mo & 1.000 & 1.000 & 1.000 & 1.000 & 1.000 & 1.000 & 1.000 & 1.000  & e   \\
 42 &  93 & $^{ 93}$Mo & 1.000 & 1.000 & 1.000 & 1.000 & 1.000 & 1.000 & 1.000 & 1.000  & n   \\
 42 &  94 & $^{ 94}$Mo & 1.000 & 1.000 & 1.000 & 1.000 & 1.000 & 1.000 & 1.000 & 1.000  & e   \\
 42 &  95 & $^{ 95}$Mo & 1.001 & 1.000 & 1.000 & 1.000 & 1.052 & 0.981 & 1.004 & 0.969  & e   \\
 42 &  96 & $^{ 96}$Mo & 1.000 & 1.000 & 1.000 & 1.000 & 1.000 & 1.000 & 1.000 & 1.000  & e   \\
 42 &  97 & $^{ 97}$Mo & 1.000 & 1.000 & 1.000 & 1.000 & 1.002 & 0.999 & 1.000 & 0.999  & e   \\
 42 &  98 & $^{ 98}$Mo & 1.000 & 1.000 & 1.000 & 1.000 & 1.000 & 1.000 & 1.000 & 1.000  & e   \\
 42 &  99 & $^{ 99}$Mo & 1.117 & 0.876 & 1.029 & 1.022 & 2.127 & 0.421 & 1.154 & 1.115  & t   \\
 42 & 100 & $^{100}$Mo & 1.000 & 1.000 & 1.000 & 1.000 & 1.006 & 0.993 & 1.002 & 1.001  & e   \\
 43 &  96 & $^{ 96}$Tc & 1.772 & 0.611 & 1.099 & 0.924 & 3.090 & 0.392 & 1.163 & 0.825  & n,$\ast$   \\
 43 &  97 & $^{ 97}$Tc & 1.009 & 0.994 & 1.002 & 0.998 & 1.125 & 0.922 & 1.018 & 0.965  & n   \\
 43 &  98 & $^{ 98}$Tc & 1.544 & 0.668 & 1.083 & 0.970 & 2.545 & 0.421 & 1.154 & 0.934  & n,$\ast$   \\
 43 &  99 & $^{ 99}$Tc & 1.011 & 0.993 & 1.002 & 0.996 & 1.237 & 0.876 & 1.030 & 0.924  & e   \\
 44 &  96 & $^{ 96}$Ru & 1.000 & 1.000 & 1.000 & 1.000 & 1.000 & 1.000 & 1.000 & 1.000  & 30   \\
 44 &  97 & $^{ 97}$Ru & 1.001 & 1.000 & 1.000 & 0.999 & 1.071 & 0.970 & 1.007 & 0.964  & n   \\
 44 &  98 & $^{ 98}$Ru & 1.000 & 1.000 & 1.000 & 1.000 & 1.001 & 0.998 & 1.000 & 1.000  & t   \\
 44 &  99 & $^{ 99}$Ru & 1.034 & 0.984 & 1.004 & 0.983 & 1.252 & 0.885 & 1.027 & 0.902  & t   \\
 44 & 100 & $^{100}$Ru & 1.000 & 1.000 & 1.000 & 1.000 & 1.006 & 0.992 & 1.002 & 1.002  & e   \\
 44 & 101 & $^{101}$Ru & 1.010 & 0.996 & 1.001 & 0.994 & 1.201 & 0.911 & 1.021 & 0.914  & e   \\
 44 & 102 & $^{102}$Ru & 1.000 & 1.000 & 1.000 & 1.000 & 1.013 & 0.984 & 1.004 & 1.003  & e   \\
 44 & 103 & $^{103}$Ru & 2.387 & 0.363 & 1.172 & 1.153 & 3.187 & 0.275 & 1.201 & 1.139  & t,$\ast$   \\
 44 & 104 & $^{104}$Ru & 1.000 & 1.000 & 1.000 & 1.000 & 1.057 & 0.933 & 1.016 & 1.014  & e   \\
 44 & 105 & $^{105}$Ru & 1.807 & 0.546 & 1.117 & 1.014 & 3.160 & 0.310 & 1.189 & 1.020  & n,$\ast$   \\
 44 & 106 & $^{106}$Ru & 1.001 & 0.999 & 1.000 & 1.000 & 1.172 & 0.814 & 1.045 & 1.047  & n   \\
 45 & 103 & $^{103}$Rh & 2.287 & 0.385 & 1.165 & 1.136 & 5.088 & 0.180 & 1.233 & 1.091  & e,$\ast$   \\
 45 & 104 & $^{104}$Rh & 1.437 & 0.697 & 1.075 & 0.998 & 4.538 & 0.224 & 1.218 & 0.983  & n,$\ast$   \\
 45 & 105 & $^{105}$Rh & 1.012 & 0.991 & 1.002 & 0.997 & 1.254 & 0.833 & 1.040 & 0.957  & n   \\
 46 & 102 & $^{102}$Pd & 1.000 & 1.000 & 1.000 & 1.000 & 1.005 & 0.995 & 1.001 & 1.001  & 30   \\
 46 & 103 & $^{103}$Pd & 1.013 & 0.993 & 1.002 & 0.994 & 1.255 & 0.869 & 1.031 & 0.916  & n   \\
 46 & 104 & $^{104}$Pd & 1.000 & 1.000 & 1.000 & 1.000 & 1.005 & 0.993 & 1.002 & 1.003  & e   \\
 46 & 105 & $^{105}$Pd & 1.000 & 1.000 & 1.000 & 1.000 & 1.083 & 0.955 & 1.011 & 0.967  & e   \\
 46 & 106 & $^{106}$Pd & 1.000 & 1.000 & 1.000 & 1.000 & 1.008 & 0.988 & 1.003 & 1.004  & e   \\
 46 & 107 & $^{107}$Pd & 1.009 & 0.994 & 1.001 & 0.997 & 1.300 & 0.813 & 1.045 & 0.946  & e   \\
 46 & 108 & $^{108}$Pd & 1.000 & 1.000 & 1.000 & 1.000 & 1.022 & 0.971 & 1.007 & 1.008  & e   \\
 46 & 109 & $^{109}$Pd & 1.012 & 0.991 & 1.002 & 0.997 & 1.512 & 0.757 & 1.059 & 0.874  & n   \\
 46 & 110 & $^{110}$Pd & 1.000 & 1.000 & 1.000 & 1.000 & 1.047 & 0.940 & 1.014 & 1.016  & e   \\
 47 & 107 & $^{107}$Ag & 1.256 & 0.753 & 1.060 & 1.057 & 3.340 & 0.262 & 1.205 & 1.142  & e,$\ast$   \\
 47 & 108 & $^{108}$Ag & 1.258 & 0.784 & 1.052 & 1.014 & 3.909 & 0.245 & 1.211 & 1.044  & n,$\ast$   \\
 47 & 109 & $^{109}$Ag & 1.273 & 0.944 & 1.013 & 0.832 & 3.342 & 0.727 & 1.067 & 0.411  & e   \\
 47 & 110 & $^{110}$Ag & 2.743 & 0.332 & 1.182 & 1.099 & 4.711 & 0.188 & 1.231 & 1.131  & t,$\ast$   \\
 47 & 111 & $^{111}$Ag & 1.610 & 0.542 & 1.118 & 1.145 & 3.989 & 0.214 & 1.222 & 1.173  & n,$\ast$   \\
 48 & 106 & $^{106}$Cd & 1.000 & 1.000 & 1.000 & 1.000 & 1.002 & 0.997 & 1.001 & 1.001  & e   \\
 48 & 107 & $^{107}$Cd & 1.002 & 0.999 & 1.000 & 1.000 & 1.132 & 0.906 & 1.022 & 0.976  & n   \\
 48 & 108 & $^{108}$Cd & 1.000 & 1.000 & 1.000 & 1.000 & 1.002 & 0.997 & 1.001 & 1.001  & e   \\
 48 & 109 & $^{109}$Cd & 1.047 & 0.968 & 1.007 & 0.986 & 1.287 & 0.824 & 1.042 & 0.943  & n   \\
 48 & 110 & $^{110}$Cd & 1.000 & 1.000 & 1.000 & 1.000 & 1.001 & 0.998 & 1.000 & 1.001  & e   \\
 48 & 111 & $^{111}$Cd & 1.001 & 0.999 & 1.000 & 1.000 & 1.236 & 0.794 & 1.050 & 1.019  & e   \\
 48 & 112 & $^{112}$Cd & 1.000 & 1.000 & 1.000 & 1.000 & 1.002 & 0.996 & 1.001 & 1.001  & e   \\
 48 & 113 & $^{113}$Cd & 1.001 & 0.999 & 1.000 & 1.000 & 1.358 & 0.695 & 1.076 & 1.060  & e   \\
 48 & 114 & $^{114}$Cd & 1.000 & 1.000 & 1.000 & 1.000 & 1.005 & 0.993 & 1.002 & 1.003  & e   \\
 48 & 115 & $^{115}$Cd & 1.015 & 0.977 & 1.005 & 1.008 & 1.870 & 0.476 & 1.137 & 1.123  & t   \\
 48 & 116 & $^{116}$Cd & 1.000 & 1.000 & 1.000 & 1.000 & 1.008 & 0.988 & 1.003 & 1.004  & e   \\
 49 & 113 & $^{113}$In & 1.000 & 1.000 & 1.000 & 1.000 & 1.002 & 0.999 & 1.000 & 1.000  & t   \\
 49 & 114 & $^{114}$In & 1.008 & 0.990 & 1.002 & 1.001 & 1.611 & 0.589 & 1.105 & 1.054  & t   \\
 49 & 115 & $^{115}$In & 1.000 & 1.000 & 1.000 & 1.000 & 1.003 & 0.997 & 1.001 & 1.000  & e   \\
 50 & 112 & $^{112}$Sn & 1.000 & 1.000 & 1.000 & 1.000 & 1.000 & 1.000 & 1.000 & 1.000  & e   \\
 50 & 114 & $^{114}$Sn & 1.000 & 1.000 & 1.000 & 1.000 & 1.000 & 1.000 & 1.000 & 1.000  & e   \\
 50 & 115 & $^{115}$Sn & 1.000 & 1.000 & 1.000 & 1.000 & 1.007 & 0.991 & 1.002 & 1.002  & e   \\
 50 & 116 & $^{116}$Sn & 1.000 & 1.000 & 1.000 & 1.000 & 1.000 & 1.000 & 1.000 & 1.000  & e   \\
 50 & 117 & $^{117}$Sn & 1.010 & 0.991 & 1.002 & 0.999 & 1.394 & 0.686 & 1.078 & 1.046  & e   \\
 50 & 118 & $^{118}$Sn & 1.000 & 1.000 & 1.000 & 1.000 & 1.000 & 1.000 & 1.000 & 1.000  & e   \\
 50 & 119 & $^{119}$Sn & 2.205 & 0.390 & 1.164 & 1.162 & 4.444 & 0.151 & 1.244 & 1.493  & e,$\ast$   \\
 50 & 120 & $^{120}$Sn & 1.000 & 1.000 & 1.000 & 1.000 & 1.000 & 1.000 & 1.000 & 1.000  & e   \\
 50 & 121 & $^{121}$Sn & 3.498 & 0.266 & 1.204 & 1.074 & 4.010 & 0.242 & 1.212 & 1.032  & t,$\ast$   \\
 50 & 122 & $^{122}$Sn & 1.000 & 1.000 & 1.000 & 1.000 & 1.000 & 1.000 & 1.000 & 1.000  & e   \\
 50 & 123 & $^{123}$Sn & 1.148 & 0.875 & 1.030 & 0.996 & 1.271 & 0.795 & 1.050 & 0.990  & n   \\
 50 & 124 & $^{124}$Sn & 1.000 & 1.000 & 1.000 & 1.000 & 1.000 & 1.000 & 1.000 & 1.000  & e   \\
 50 & 125 & $^{125}$Sn & 1.133 & 0.881 & 1.028 & 1.001 & 1.248 & 0.803 & 1.048 & 0.998  & t   \\
 50 & 126 & $^{126}$Sn & 1.000 & 1.000 & 1.000 & 1.000 & 1.000 & 1.000 & 1.000 & 1.000  & t   \\
 51 & 121 & $^{121}$Sb & 1.387 & 0.724 & 1.068 & 0.996 & 1.840 & 0.528 & 1.122 & 1.029  & 30,$\ast$   \\
 51 & 122 & $^{122}$Sb & 1.342 & 0.762 & 1.058 & 0.978 & 3.978 & 0.268 & 1.203 & 0.938  & t,$\ast$   \\
 51 & 123 & $^{123}$Sb & 1.004 & 0.998 & 1.001 & 0.999 & 1.102 & 0.938 & 1.015 & 0.968  & 30   \\
 51 & 125 & $^{125}$Sb & 1.000 & 1.000 & 1.000 & 1.000 & 1.012 & 0.994 & 1.002 & 0.995  & t   \\
 51 & 126 & $^{126}$Sb & 1.486 & 0.638 & 1.091 & 1.054 & 2.095 & 0.494 & 1.132 & 0.966  & n,$\ast$   \\
 51 & 127 & $^{127}$Sb & 1.000 & 1.000 & 1.000 & 1.000 & 1.002 & 0.999 & 1.000 & 0.999  & n   \\
 52 & 120 & $^{120}$Te & 1.000 & 1.000 & 1.000 & 1.000 & 1.004 & 0.993 & 1.002 & 1.003  & 30   \\
 52 & 122 & $^{122}$Te & 1.000 & 1.000 & 1.000 & 1.000 & 1.004 & 0.994 & 1.002 & 1.002  & e   \\
 52 & 123 & $^{123}$Te & 1.012 & 0.986 & 1.003 & 1.002 & 1.576 & 0.557 & 1.114 & 1.140  & e   \\
 52 & 124 & $^{124}$Te & 1.000 & 1.000 & 1.000 & 1.000 & 1.003 & 0.996 & 1.001 & 1.002  & e   \\
 52 & 125 & $^{125}$Te & 1.661 & 0.546 & 1.117 & 1.102 & 3.410 & 0.204 & 1.225 & 1.437  & e,$\ast$   \\
 52 & 126 & $^{126}$Te & 1.000 & 1.000 & 1.000 & 1.000 & 1.001 & 0.998 & 1.000 & 1.001  & e   \\
 52 & 127 & $^{127}$Te & 1.223 & 0.797 & 1.049 & 1.025 & 2.271 & 0.403 & 1.160 & 1.093  & n,$\ast$   \\
 52 & 128 & $^{128}$Te & 1.000 & 1.000 & 1.000 & 1.000 & 1.000 & 0.999 & 1.000 & 1.000  & e   \\
 52 & 130 & $^{130}$Te & 1.000 & 1.000 & 1.000 & 1.000 & 1.000 & 1.000 & 1.000 & 1.000  & e   \\
 53 & 127 & $^{127}$I  & 1.196 & 0.847 & 1.037 & 0.987 & 1.712 & 0.584 & 1.106 & 1.000  & e   \\
 53 & 129 & $^{129}$I  & 1.297 & 0.823 & 1.043 & 0.937 & 1.548 & 0.717 & 1.070 & 0.902  & e,$\ast$   \\
 53 & 130 & $^{130}$I  & 1.960 & 0.647 & 1.089 & 0.789 & 4.752 & 0.328 & 1.184 & 0.641  & n,$\ast$   \\
 53 & 131 & $^{131}$I  & 1.005 & 0.998 & 1.001 & 0.997 & 1.117 & 0.945 & 1.013 & 0.948  & n   \\
 54 & 124 & $^{124}$Xe & 1.000 & 1.000 & 1.000 & 1.000 & 1.060 & 0.915 & 1.020 & 1.031  & 30   \\
 54 & 126 & $^{126}$Xe & 1.000 & 1.000 & 1.000 & 1.000 & 1.039 & 0.940 & 1.014 & 1.024  & 30   \\
 54 & 128 & $^{128}$Xe & 1.000 & 1.000 & 1.000 & 1.000 & 1.020 & 0.970 & 1.007 & 1.011  & 30   \\
 54 & 129 & $^{129}$Xe & 1.537 & 0.608 & 1.100 & 1.071 & 2.820 & 0.271 & 1.202 & 1.309  & 30,$\ast$   \\
 54 & 130 & $^{130}$Xe & 1.000 & 1.000 & 1.000 & 1.000 & 1.006 & 0.990 & 1.002 & 1.003  & 30   \\
 54 & 131 & $^{131}$Xe & 1.047 & 0.965 & 1.008 & 0.989 & 1.629 & 0.564 & 1.112 & 1.089  & t   \\
 54 & 132 & $^{132}$Xe & 1.000 & 1.000 & 1.000 & 1.000 & 1.001 & 0.998 & 1.000 & 1.001  & e   \\
 54 & 133 & $^{133}$Xe & 1.001 & 0.998 & 1.000 & 1.001 & 1.185 & 0.801 & 1.048 & 1.053  & t   \\
 54 & 134 & $^{134}$Xe & 1.000 & 1.000 & 1.000 & 1.000 & 1.000 & 1.000 & 1.000 & 1.000  & 30   \\
 54 & 136 & $^{136}$Xe & 1.000 & 1.000 & 1.000 & 1.000 & 1.000 & 1.000 & 1.000 & 1.000  & e   \\
 55 & 133 & $^{133}$Cs & 1.054 & 0.966 & 1.008 & 0.983 & 1.379 & 0.818 & 1.044 & 0.886  & e   \\
 55 & 134 & $^{134}$Cs & 1.973 & 0.510 & 1.127 & 0.993 & 3.305 & 0.305 & 1.191 & 0.991  & t,$\ast$   \\
 55 & 135 & $^{135}$Cs & 1.000 & 1.000 & 1.000 & 1.000 & 1.049 & 0.972 & 1.007 & 0.981  & 30   \\
 55 & 136 & $^{136}$Cs & 1.043 & 0.832 & 1.040 & 1.152 & 1.359 & 0.610 & 1.099 & 1.206  & n,$\ast$   \\
 55 & 137 & $^{137}$Cs & 1.000 & 1.000 & 1.000 & 1.000 & 1.003 & 0.999 & 1.000 & 0.999  & n   \\
 56 & 130 & $^{130}$Ba & 1.000 & 1.000 & 1.000 & 1.000 & 1.065 & 0.907 & 1.022 & 1.035  & 30   \\
 56 & 132 & $^{132}$Ba & 1.000 & 1.000 & 1.000 & 1.000 & 1.015 & 0.980 & 1.005 & 1.005  & 30   \\
 56 & 134 & $^{134}$Ba & 1.000 & 1.000 & 1.000 & 1.000 & 1.003 & 0.996 & 1.001 & 1.001  & e   \\
 56 & 135 & $^{135}$Ba & 1.001 & 0.999 & 1.000 & 1.000 & 1.141 & 0.818 & 1.044 & 1.071  & e   \\
 56 & 136 & $^{136}$Ba & 1.000 & 1.000 & 1.000 & 1.000 & 1.000 & 1.000 & 1.000 & 1.000  & e   \\
 56 & 137 & $^{137}$Ba & 1.000 & 1.000 & 1.000 & 1.000 & 1.016 & 0.994 & 1.001 & 0.991  & e   \\
 56 & 138 & $^{138}$Ba & 1.000 & 1.000 & 1.000 & 1.000 & 1.000 & 1.000 & 1.000 & 1.000  & e   \\
 57 & 138 & $^{138}$La & 1.070 & 0.949 & 1.012 & 0.984 & 1.550 & 0.768 & 1.056 & 0.839  & t   \\
 57 & 139 & $^{139}$La & 1.003 & 0.999 & 1.000 & 0.998 & 1.094 & 0.955 & 1.010 & 0.957  & e   \\
 57 & 140 & $^{140}$La & 2.439 & 0.485 & 1.135 & 0.846 & 5.104 & 0.238 & 1.213 & 0.822  & n,$\ast$   \\
 58 & 132 & $^{132}$Ce & 1.000 & 1.000 & 1.000 & 1.000 & 1.086 & 0.882 & 1.028 & 1.044  & t   \\
 58 & 133 & $^{133}$Ce & 2.482 & 0.294 & 1.195 & 1.370 & 5.223 & 0.133 & 1.250 & 1.440  & t,$\ast$   \\
 58 & 134 & $^{134}$Ce & 1.000 & 1.000 & 1.000 & 1.000 & 1.030 & 0.954 & 1.011 & 1.018  & t   \\
 58 & 135 & $^{135}$Ce & 1.127 & 0.874 & 1.030 & 1.015 & 1.827 & 0.489 & 1.134 & 1.119  & t   \\
 58 & 136 & $^{136}$Ce & 1.000 & 1.000 & 1.000 & 1.000 & 1.005 & 0.992 & 1.002 & 1.003  & 30   \\
 58 & 137 & $^{137}$Ce & 1.003 & 0.997 & 1.001 & 1.000 & 1.203 & 0.747 & 1.062 & 1.113  & t   \\
 58 & 138 & $^{138}$Ce & 1.000 & 1.000 & 1.000 & 1.000 & 1.000 & 1.000 & 1.000 & 1.000  & 30   \\
 58 & 139 & $^{139}$Ce & 1.000 & 1.000 & 1.000 & 1.000 & 1.021 & 0.991 & 1.002 & 0.988  & t   \\
 58 & 140 & $^{140}$Ce & 1.000 & 1.000 & 1.000 & 1.000 & 1.000 & 1.000 & 1.000 & 1.000  & e   \\
 58 & 141 & $^{141}$Ce & 1.000 & 1.000 & 1.000 & 1.000 & 1.000 & 1.000 & 1.000 & 1.000  & t   \\
 58 & 142 & $^{142}$Ce & 1.000 & 1.000 & 1.000 & 1.000 & 1.002 & 0.997 & 1.001 & 1.001  & 30   \\
 58 & 143 & $^{143}$Ce & 2.431 & 0.358 & 1.174 & 1.149 & 3.464 & 0.207 & 1.224 & 1.395  & n,$\ast$   \\
 59 & 141 & $^{141}$Pr & 1.010 & 0.991 & 1.002 & 0.998 & 1.217 & 0.829 & 1.041 & 0.992  & e   \\
 59 & 142 & $^{142}$Pr & 4.406 & 0.203 & 1.226 & 1.120 & 8.179 & 0.122 & 1.254 & 1.002  & t,$\ast$   \\
 59 & 143 & $^{143}$Pr & 1.111 & 0.933 & 1.016 & 0.965 & 1.375 & 0.800 & 1.048 & 0.910  & t   \\
 60 & 142 & $^{142}$Nd & 1.000 & 1.000 & 1.000 & 1.000 & 1.000 & 1.000 & 1.000 & 1.000  & e   \\
 60 & 143 & $^{143}$Nd & 1.000 & 1.000 & 1.000 & 1.000 & 1.000 & 1.000 & 1.000 & 1.000  & e   \\
 60 & 144 & $^{144}$Nd & 1.000 & 1.000 & 1.000 & 1.000 & 1.001 & 0.998 & 1.000 & 1.001  & e   \\
 60 & 145 & $^{145}$Nd & 1.120 & 0.943 & 1.013 & 0.947 & 1.520 & 0.793 & 1.050 & 0.830  & e   \\
 60 & 146 & $^{146}$Nd & 1.000 & 1.000 & 1.000 & 1.000 & 1.017 & 0.977 & 1.005 & 1.006  & e   \\
 60 & 147 & $^{147}$Nd & 1.270 & 0.807 & 1.047 & 0.976 & 2.112 & 0.487 & 1.134 & 0.971  & t,$\ast$   \\
 60 & 148 & $^{148}$Nd & 1.000 & 1.000 & 1.000 & 1.000 & 1.116 & 0.884 & 1.027 & 1.013  & e   \\
 60 & 150 & $^{150}$Nd & 1.065 & 0.926 & 1.017 & 1.014 & 2.061 & 0.423 & 1.154 & 1.147  & e   \\
 61 & 147 & $^{147}$Pm & 1.057 & 0.963 & 1.009 & 0.983 & 1.672 & 0.660 & 1.085 & 0.907  & 30   \\
 61 & 148 & $^{148}$Pm & 1.165 & 0.854 & 1.035 & 1.005 & 4.622 & 0.200 & 1.226 & 1.081  & t   \\
 61 & 149 & $^{149}$Pm & 1.019 & 0.988 & 1.003 & 0.993 & 1.464 & 0.786 & 1.052 & 0.869  & t   \\
 62 & 144 & $^{144}$Sm & 1.000 & 1.000 & 1.000 & 1.000 & 1.000 & 1.000 & 1.000 & 1.000  & e   \\
 62 & 147 & $^{147}$Sm & 1.014 & 0.992 & 1.002 & 0.995 & 1.208 & 0.890 & 1.026 & 0.931  & e   \\
 62 & 148 & $^{148}$Sm & 1.000 & 1.000 & 1.000 & 1.000 & 1.005 & 0.992 & 1.002 & 1.003  & e   \\
 62 & 149 & $^{149}$Sm & 1.354 & 0.803 & 1.048 & 0.919 & 1.634 & 0.682 & 1.079 & 0.897  & e,$\ast$   \\
 62 & 150 & $^{150}$Sm & 1.000 & 1.000 & 1.000 & 1.000 & 1.078 & 0.913 & 1.021 & 1.017  & e   \\
 62 & 151 & $^{151}$Sm & 1.945 & 0.585 & 1.106 & 0.879 & 4.595 & 0.244 & 1.211 & 0.892  & e,$\ast$  \\
 62 & 152 & $^{152}$Sm & 1.086 & 0.890 & 1.026 & 1.035 & 2.185 & 0.363 & 1.172 & 1.259  & e   \\
 62 & 153 & $^{153}$Sm & 3.321 & 0.243 & 1.212 & 1.241 & 8.509 & 0.086 & 1.267 & 1.364  & t,$\ast$   \\
 62 & 154 & $^{154}$Sm & 1.326 & 0.670 & 1.083 & 1.126 & 3.129 & 0.219 & 1.220 & 1.459  & e,$\ast$   \\
 63 & 151 & $^{151}$Eu & 1.656 & 0.582 & 1.107 & 1.037 & 2.506 & 0.379 & 1.167 & 1.052  & e,$\ast$   \\
 63 & 152 & $^{152}$Eu & 1.557 & 0.706 & 1.073 & 0.910 & 8.333 & 0.165 & 1.239 & 0.730  & t,$\ast$   \\
 63 & 153 & $^{153}$Eu & 1.158 & 0.884 & 1.027 & 0.976 & 2.650 & 0.441 & 1.148 & 0.856  & e   \\
 63 & 154 & $^{154}$Eu & 1.415 & 0.752 & 1.061 & 0.940 & 5.848 & 0.223 & 1.219 & 0.768  & 30,$\ast$   \\
 63 & 155 & $^{155}$Eu & 1.138 & 0.893 & 1.025 & 0.985 & 2.332 & 0.473 & 1.138 & 0.906  & 30   \\
 63 & 156 & $^{156}$Eu & 4.159 & 0.170 & 1.237 & 1.413 & 19.94 & 0.033 & 1.287 & 1.523  & n,$\ast$   \\
 64 & 152 & $^{152}$Gd & 1.000 & 1.000 & 1.000 & 1.000 & 1.069 & 0.908 & 1.022 & 1.030  & e   \\
 64 & 153 & $^{153}$Gd & 1.681 & 0.580 & 1.107 & 1.026 & 6.064 & 0.162 & 1.240 & 1.020  & t,$\ast$   \\
 64 & 154 & $^{154}$Gd & 1.083 & 0.880 & 1.029 & 1.050 & 2.163 & 0.352 & 1.176 & 1.312  & e   \\
 64 & 155 & $^{155}$Gd & 1.499 & 0.662 & 1.085 & 1.008 & 5.528 & 0.180 & 1.233 & 1.006  & e,$\ast$   \\
 64 & 156 & $^{156}$Gd & 1.258 & 0.703 & 1.074 & 1.131 & 2.898 & 0.239 & 1.213 & 1.442  & e,$\ast$   \\
 64 & 157 & $^{157}$Gd & 1.497 & 0.685 & 1.078 & 0.975 & 3.930 & 0.256 & 1.207 & 0.993  & e,$\ast$   \\
 64 & 158 & $^{158}$Gd & 1.355 & 0.652 & 1.087 & 1.132 & 3.209 & 0.211 & 1.223 & 1.478  & e,$\ast$   \\
 64 & 159 & $^{159}$Gd & 1.527 & 0.675 & 1.081 & 0.971 & 4.296 & 0.250 & 1.209 & 0.932  & n,$\ast$   \\
 64 & 160 & $^{160}$Gd & 1.409 & 0.629 & 1.094 & 1.128 & 3.375 & 0.196 & 1.228 & 1.509  & n,$\ast$   \\
 65 & 157 & $^{157}$Tb & 1.214 & 0.817 & 1.044 & 1.008 & 2.282 & 0.413 & 1.157 & 1.060  & n,$\ast$   \\
 65 & 158 & $^{158}$Tb & 1.339 & 0.743 & 1.063 & 1.005 & 3.227 & 0.321 & 1.186 & 0.965  & n,$\ast$   \\
 65 & 159 & $^{159}$Tb & 1.238 & 0.785 & 1.052 & 1.029 & 2.327 & 0.398 & 1.161 & 1.079  & e,$\ast$   \\
 65 & 160 & $^{160}$Tb & 1.382 & 0.747 & 1.062 & 0.969 & 3.993 & 0.302 & 1.192 & 0.829  & t,$\ast$   \\
 65 & 161 & $^{161}$Tb & 1.254 & 0.793 & 1.050 & 1.006 & 2.333 & 0.402 & 1.160 & 1.067  & n,$\ast$   \\
 66 & 156 & $^{156}$Dy & 1.051 & 0.922 & 1.018 & 1.033 & 1.952 & 0.401 & 1.160 & 1.278  & 30   \\
 66 & 158 & $^{158}$Dy & 1.185 & 0.820 & 1.043 & 1.030 & 2.628 & 0.290 & 1.196 & 1.311  & t,$\ast$   \\
 66 & 159 & $^{159}$Dy & 1.256 & 0.790 & 1.051 & 1.008 & 2.908 & 0.338 & 1.180 & 1.017  & n,$\ast$   \\
 66 & 160 & $^{160}$Dy & 1.278 & 0.696 & 1.075 & 1.125 & 2.958 & 0.234 & 1.215 & 1.442  & e,$\ast$   \\
 66 & 161 & $^{161}$Dy & 1.913 & 0.530 & 1.122 & 0.987 & 4.433 & 0.232 & 1.215 & 0.972  & e,$\ast$   \\
 66 & 162 & $^{162}$Dy & 1.341 & 0.636 & 1.092 & 1.173 & 3.162 & 0.212 & 1.222 & 1.493  & e,$\ast$   \\
 66 & 163 & $^{163}$Dy & 1.122 & 0.896 & 1.025 & 0.994 & 1.965 & 0.534 & 1.120 & 0.953  & e   \\
 66 & 164 & $^{164}$Dy & 1.436 & 0.596 & 1.103 & 1.168 & 3.459 & 0.189 & 1.230 & 1.528  & e,$\ast$   \\
 67 & 163 & $^{163}$Ho & 1.046 & 0.959 & 1.009 & 0.997 & 1.521 & 0.676 & 1.081 & 0.973  & 30   \\
 67 & 165 & $^{165}$Ho & 1.055 & 0.950 & 1.012 & 0.999 & 1.539 & 0.670 & 1.082 & 0.970  & e   \\
 67 & 166 & $^{166}$Ho & 14.53 & 0.042 & 1.284 & 1.648 & 26.69 & 0.023 & 1.291 & 1.602  & n,$\ast$   \\
 68 & 162 & $^{162}$Er & 1.167 & 0.784 & 1.053 & 1.094 & 2.547 & 0.291 & 1.195 & 1.347  & 30,$\ast$   \\
 68 & 164 & $^{164}$Er & 1.238 & 0.745 & 1.062 & 1.084 & 2.815 & 0.268 & 1.203 & 1.328  & 30,$\ast$   \\
 68 & 166 & $^{166}$Er & 1.342 & 0.636 & 1.092 & 1.172 & 3.168 & 0.214 & 1.221 & 1.472  & e,$\ast$   \\
 68 & 167 & $^{167}$Er & 1.093 & 0.930 & 1.016 & 0.983 & 1.770 & 0.598 & 1.102 & 0.944  & e   \\
 68 & 168 & $^{168}$Er & 1.351 & 0.653 & 1.087 & 1.133 & 3.190 & 0.211 & 1.223 & 1.485  & e,$\ast$   \\
 68 & 169 & $^{169}$Er & 1.638 & 0.611 & 1.099 & 0.999 & 5.473 & 0.172 & 1.236 & 1.064  & t,$\ast$   \\
 68 & 170 & $^{170}$Er & 1.366 & 0.651 & 1.088 & 1.125 & 3.236 & 0.205 & 1.225 & 1.506  & e,$\ast$   \\
 69 & 169 & $^{169}$Tm & 2.609 & 0.304 & 1.191 & 1.260 & 4.582 & 0.159 & 1.241 & 1.375  & e,$\ast$   \\
 69 & 170 & $^{170}$Tm & 1.529 & 0.618 & 1.097 & 1.058 & 4.041 & 0.237 & 1.214 & 1.046  & t,$\ast$   \\
 69 & 171 & $^{171}$Tm & 2.808 & 0.263 & 1.205 & 1.356 & 4.573 & 0.144 & 1.246 & 1.521  & t,$\ast$   \\
 69 & 172 & $^{172}$Tm & 1.189 & 0.841 & 1.038 & 1.000 & 2.015 & 0.486 & 1.135 & 1.021  & n   \\
 70 & 168 & $^{168}$Yb & 1.269 & 0.713 & 1.071 & 1.105 & 2.930 & 0.251 & 1.209 & 1.361  & 30,$\ast$   \\
 70 & 170 & $^{170}$Yb & 1.302 & 0.667 & 1.083 & 1.151 & 3.034 & 0.230 & 1.216 & 1.433  & e,$\ast$   \\
 70 & 171 & $^{171}$Yb & 1.700 & 0.551 & 1.116 & 1.067 & 6.731 & 0.129 & 1.252 & 1.151  & e,$\ast$   \\
 70 & 172 & $^{172}$Yb & 1.364 & 0.624 & 1.095 & 1.174 & 3.233 & 0.207 & 1.224 & 1.497  & e,$\ast$   \\
 70 & 173 & $^{173}$Yb & 1.101 & 0.916 & 1.020 & 0.991 & 1.768 & 0.578 & 1.108 & 0.979  & e   \\
 70 & 174 & $^{174}$Yb & 1.392 & 0.633 & 1.092 & 1.134 & 3.320 & 0.198 & 1.227 & 1.519  & e,$\ast$   \\
 70 & 175 & $^{175}$Yb & 1.039 & 0.968 & 1.008 & 0.994 & 1.500 & 0.715 & 1.071 & 0.933  & t   \\
 70 & 176 & $^{176}$Yb & 1.325 & 0.673 & 1.082 & 1.121 & 3.104 & 0.214 & 1.222 & 1.508  & e,$\ast$   \\
 71 & 175 & $^{175}$Lu & 1.028 & 0.975 & 1.006 & 0.998 & 1.427 & 0.728 & 1.067 & 0.963  & e   \\
 71 & 176 & $^{176}$Lu\tablenotemark{b} & 1.006 & 0.995 & 1.001 & 0.999 & 1.286 & 0.830 & 1.041 & 0.937  & e   \\
 71 & 177 & $^{177}$Lu & 1.030 & 0.975 & 1.006 & 0.996 & 1.576 & 0.695 & 1.076 & 0.912  & n   \\
 72 & 174 & $^{174}$Hf & 1.241 & 0.730 & 1.067 & 1.104 & 2.828 & 0.260 & 1.206 & 1.359  & 30,$\ast$   \\
 72 & 176 & $^{176}$Hf & 1.264 & 0.719 & 1.069 & 1.101 & 2.905 & 0.244 & 1.211 & 1.408  & e,$\ast$   \\
 72 & 177 & $^{177}$Hf & 1.029 & 0.973 & 1.006 & 0.999 & 1.418 & 0.724 & 1.068 & 0.973  & e   \\
 72 & 178 & $^{178}$Hf & 1.224 & 0.740 & 1.064 & 1.104 & 2.759 & 0.257 & 1.207 & 1.413  & e,$\ast$   \\
 72 & 179 & $^{179}$Hf & 1.021 & 0.986 & 1.003 & 0.994 & 1.395 & 0.763 & 1.058 & 0.939  & e   \\
 72 & 180 & $^{180}$Hf & 1.223 & 0.750 & 1.061 & 1.090 & 2.752 & 0.258 & 1.207 & 1.410  & e,$\ast$   \\
 72 & 181 & $^{181}$Hf & 1.552 & 0.605 & 1.100 & 1.065 & 3.611 & 0.215 & 1.221 & 1.285  & t,$\ast$   \\
 72 & 182 & $^{182}$Hf & 1.192 & 0.780 & 1.053 & 1.075 & 2.636 & 0.268 & 1.203 & 1.418  & 30,$\ast$   \\
 73 & 179 & $^{179}$Ta & 1.468 & 0.677 & 1.081 & 1.007 & 2.365 & 0.434 & 1.150 & 0.974  & t,$\ast$   \\
 73 & 180 & $^{180}$Ta\tablenotemark{b} & 2.037 & 0.437 & 1.149 & 1.122 & 7.168 & 0.109 & 1.259 & 1.281  & e,$\ast$   \\
 73 & 181 & $^{181}$Ta & 2.038 & 0.478 & 1.137 & 1.028 & 2.660 & 0.378 & 1.168 & 0.995  & e,$\ast$   \\
 73 & 182 & $^{182}$Ta & 2.014 & 0.492 & 1.133 & 1.008 & 3.987 & 0.285 & 1.197 & 0.879  & t,$\ast$   \\
 73 & 183 & $^{183}$Ta & 1.124 & 0.909 & 1.021 & 0.979 & 1.952 & 0.547 & 1.117 & 0.938  & n   \\
 74 & 180 & $^{180}$W  & 1.159 & 0.805 & 1.047 & 1.072 & 2.506 & 0.301 & 1.192 & 1.324  & 30,$\ast$   \\
 74 & 182 & $^{182}$W  & 1.178 & 0.793 & 1.050 & 1.070 & 2.580 & 0.284 & 1.198 & 1.366  & e,$\ast$   \\
 74 & 183 & $^{183}$W  & 1.541 & 0.599 & 1.102 & 1.083 & 3.827 & 0.206 & 1.224 & 1.265  & e,$\ast$   \\
 74 & 184 & $^{184}$W  & 1.123 & 0.841 & 1.038 & 1.059 & 2.342 & 0.320 & 1.186 & 1.335  & 30   \\
 74 & 185 & $^{185}$W  & 1.454 & 0.751 & 1.061 & 0.916 & 3.217 & 0.315 & 1.188 & 0.989  & 30,$\ast$   \\
 74 & 186 & $^{186}$W  & 1.084 & 0.905 & 1.022 & 1.019 & 2.145 & 0.380 & 1.167 & 1.227  & 30   \\
 74 & 187 & $^{187}$W  & 1.121 & 0.897 & 1.024 & 0.994 & 2.004 & 0.513 & 1.127 & 0.973  & n   \\
 75 & 185 & $^{185}$Re & 1.020 & 0.981 & 1.004 & 0.999 & 1.351 & 0.744 & 1.063 & 0.996  & 30   \\
 75 & 186 & $^{186}$Re & 1.396 & 0.689 & 1.077 & 1.039 & 5.199 & 0.168 & 1.238 & 1.145  & t ,$\ast$  \\
 75 & 187 & $^{187}$Re & 1.017 & 0.987 & 1.003 & 0.996 & 1.432 & 0.723 & 1.068 & 0.966  & 30   \\
 75 & 188 & $^{188}$Re & 1.237 & 0.795 & 1.050 & 1.017 & 3.349 & 0.263 & 1.205 & 1.134  & n,$\ast$   \\
 75 & 189 & $^{189}$Re & 1.036 & 0.964 & 1.008 & 1.001 & 1.640 & 0.620 & 1.096 & 0.984  & n   \\
 76 & 184 & $^{184}$Os & 1.092 & 0.884 & 1.027 & 1.036 & 2.194 & 0.362 & 1.173 & 1.260  & 30   \\
 76 & 186 & $^{186}$Os & 1.052 & 0.917 & 1.020 & 1.037 & 1.940 & 0.410 & 1.158 & 1.257  & e   \\
 76 & 187 & $^{187}$Os & 3.016 & 0.278 & 1.200 & 1.193 & 7.184 & 0.104 & 1.260 & 1.333  & e,$\ast$   \\
 76 & 188 & $^{188}$Os & 1.028 & 0.957 & 1.010 & 1.016 & 1.746 & 0.474 & 1.138 & 1.209  & e   \\
 76 & 189 & $^{189}$Os & 2.356 & 0.384 & 1.166 & 1.105 & 5.402 & 0.153 & 1.243 & 1.209  & e,$\ast$   \\
 76 & 190 & $^{190}$Os & 1.010 & 0.985 & 1.004 & 1.006 & 1.500 & 0.580 & 1.107 & 1.150  & 30   \\
 76 & 191 & $^{191}$Os & 1.060 & 0.967 & 1.008 & 0.976 & 1.639 & 0.750 & 1.061 & 0.813  & t   \\
 76 & 192 & $^{192}$Os & 1.005 & 0.993 & 1.002 & 1.002 & 1.401 & 0.658 & 1.086 & 1.085  & 30   \\
 76 & 193 & $^{193}$Os & 1.290 & 0.844 & 1.037 & 0.918 & 2.281 & 0.486 & 1.134 & 0.901  & n   \\
 76 & 194 & $^{194}$Os & 1.003 & 0.995 & 1.001 & 1.001 & 1.333 & 0.691 & 1.077 & 1.085  & n   \\
 77 & 191 & $^{191}$Ir & 1.065 & 0.949 & 1.012 & 0.990 & 2.015 & 0.473 & 1.139 & 1.050  & e   \\
 77 & 192 & $^{192}$Ir & 3.065 & 0.331 & 1.183 & 0.986 & 6.000 & 0.183 & 1.232 & 0.910  & t,$\ast$   \\
 77 & 193 & $^{193}$Ir & 1.268 & 0.747 & 1.062 & 1.056 & 2.794 & 0.299 & 1.193 & 1.197  & e,$\ast$   \\
 77 & 194 & $^{194}$Ir & 1.349 & 0.789 & 1.051 & 0.940 & 4.830 & 0.222 & 1.219 & 0.931  & n,$\ast$   \\
 78 & 190 & $^{190}$Pt & 1.000 & 1.000 & 1.000 & 1.000 & 1.128 & 0.845 & 1.037 & 1.050  & 30   \\
 78 & 192 & $^{192}$Pt & 1.000 & 1.000 & 1.000 & 1.000 & 1.099 & 0.869 & 1.031 & 1.048  & 30   \\
 78 & 193 & $^{193}$Pt & 4.881 & 0.156 & 1.242 & 1.312 & 8.241 & 0.080 & 1.270 & 1.517  & t,$\ast$   \\
 78 & 194 & $^{194}$Pt & 1.000 & 1.000 & 1.000 & 1.000 & 1.085 & 0.890 & 1.026 & 1.035  & t   \\
 78 & 195 & $^{195}$Pt & 1.121 & 0.888 & 1.026 & 1.004 & 3.119 & 0.288 & 1.197 & 1.112  & t   \\
 78 & 196 & $^{196}$Pt & 1.000 & 1.000 & 1.000 & 1.000 & 1.060 & 0.925 & 1.018 & 1.021  & 30   \\
 78 & 197 & $^{197}$Pt & 1.783 & 0.536 & 1.120 & 1.046 & 4.332 & 0.193 & 1.229 & 1.197  & n,$\ast$   \\
 78 & 198 & $^{198}$Pt & 1.000 & 1.000 & 1.000 & 1.000 & 1.031 & 0.964 & 1.008 & 1.006  & 30   \\
 79 & 195 & $^{195}$Au & 1.065 & 0.962 & 1.009 & 0.976 & 1.403 & 0.750 & 1.061 & 0.950  & n   \\
 79 & 196 & $^{196}$Au & 1.677 & 0.631 & 1.093 & 0.945 & 3.512 & 0.298 & 1.193 & 0.956  & n,$\ast$   \\
 79 & 197 & $^{197}$Au & 1.038 & 0.983 & 1.004 & 0.980 & 1.294 & 0.841 & 1.038 & 0.919  & e   \\
 79 & 198 & $^{198}$Au & 1.108 & 0.944 & 1.013 & 0.956 & 1.858 & 0.665 & 1.084 & 0.809  & t   \\
 79 & 199 & $^{199}$Au & 1.038 & 0.988 & 1.003 & 0.975 & 1.246 & 0.897 & 1.024 & 0.895  & n   \\
 80 & 196 & $^{196}$Hg & 1.000 & 1.000 & 1.000 & 1.000 & 1.024 & 0.963 & 1.009 & 1.014  & 30   \\
 80 & 198 & $^{198}$Hg & 1.000 & 1.000 & 1.000 & 1.000 & 1.029 & 0.961 & 1.009 & 1.011  & e   \\
 80 & 199 & $^{199}$Hg & 1.017 & 0.978 & 1.005 & 1.005 & 1.611 & 0.590 & 1.104 & 1.052  & e   \\
 80 & 200 & $^{200}$Hg & 1.000 & 1.000 & 1.000 & 1.000 & 1.050 & 0.941 & 1.014 & 1.012  & e   \\
 80 & 201 & $^{201}$Hg & 2.442 & 0.449 & 1.146 & 0.911 & 3.332 & 0.282 & 1.199 & 1.064  & e,$\ast$   \\
 80 & 202 & $^{202}$Hg & 1.000 & 1.000 & 1.000 & 1.000 & 1.021 & 0.979 & 1.005 & 1.000  & 30   \\
 80 & 203 & $^{203}$Hg & 1.383 & 0.832 & 1.040 & 0.869 & 1.731 & 0.733 & 1.066 & 0.789  & t,$\ast$   \\
 80 & 204 & $^{204}$Hg & 1.000 & 1.000 & 1.000 & 1.000 & 1.021 & 0.982 & 1.004 & 0.997  & 30   \\
 81 & 203 & $^{203}$Tl & 1.000 & 1.000 & 1.000 & 1.000 & 1.062 & 0.949 & 1.012 & 0.992  & e   \\
 81 & 204 & $^{204}$Tl & 1.007 & 0.998 & 1.001 & 0.995 & 1.170 & 0.932 & 1.016 & 0.917  & t   \\
 81 & 205 & $^{205}$Tl & 1.002 & 0.998 & 1.000 & 0.999 & 1.158 & 0.899 & 1.024 & 0.960  & e   \\
 82 & 204 & $^{204}$Pb & 1.000 & 1.000 & 1.000 & 1.000 & 1.000 & 1.000 & 1.000 & 1.000  & e   \\
 82 & 205 & $^{205}$Pb & 1.309 & 0.824 & 1.042 & 0.927 & 1.350 & 0.836 & 1.039 & 0.886  & t,$\ast$   \\
 82 & 206 & $^{206}$Pb & 1.000 & 1.000 & 1.000 & 1.000 & 1.000 & 1.000 & 1.000 & 1.000  & e   \\
 82 & 207 & $^{207}$Pb & 1.000 & 1.000 & 1.000 & 1.000 & 1.002 & 0.998 & 1.001 & 1.000  & e   \\
 82 & 208 & $^{208}$Pb & 1.000 & 1.000 & 1.000 & 1.000 & 1.000 & 1.000 & 1.000 & 1.000  & e   \\
 83 & 209 & $^{209}$Bi & 1.000 & 1.000 & 1.000 & 1.000 & 1.000 & 1.000 & 1.000 & 1.000  & e   \\
 83 & 210 & $^{210}$Bi & 1.072 & 0.981 & 1.004 & 0.951 & 1.514 & 0.625 & 1.095 & 1.056  & t   \\
\enddata
%% Text for table notes should follow after the \enddata but before
%% the \end{deluxetable}. Make sure there is at least one \tablenotemark
%% in the table for each \tablenotetext.
%\tablecomments{Table \ref{tab:results} is published in its entirety in the
%electronic edition of the {\it Astrophysical Journal}.  A portion is
%shown here for guidance regarding its form and content.}
\tablenotetext{a}{The marks appearing in the comments refer to KADoNiS v0.3 \citep{KADONIS}: n (not present), t (only theoretical estimate), 30 (only 30 keV MACS), e (measured in the relevant energy range), $*$ ($X_\mathrm{lower}^{30}<0.8$, appearing in Figure \ref{fig:surprise}).}
\tablenotetext{b}{$X$ and $f_\mathrm{SEF}$ given for the g.s., not the isomeric state, and assuming thermal equilibration of g.s.\ and excited states.}
\end{deluxetable}


\begin{thebibliography}{99}

\bibitem[Achterberg et al.(2006)]{Ach06}
Achterberg, E., Capurro, O. A., Marti, G. V., Vanin, V. R., \& Castro, R. M. 2006,
Nucl.\ Data Sheets, 107, 1

\bibitem[Bao et al.(2000)]{Bao00}
Bao, Z. Y., Beer, H., K\"appeler, F., Voss, F., Wisshak, K., \&
Rauscher, T. 2000,
At.\ Data Nucl.\ Data Tables, 76, 70

\bibitem[Basu, Mukherjee, \& Sonzogni(2010)]{Basu10}
Basu, S. K., Mukherjee, G., \& Sonzogni, A. A. 2010,
Nucl.\ Data Sheets, 111, 2555
%
\bibitem[Basunia(2009)]{Bas10}
Basunia, M. S. 2009,
Nucl.\ Data Sheets, 110, 999
%
\bibitem[Belic et al.(1999)]{Bel99}
Belic, D., et al.\ 1999,
%C.~Arlandini, J.~Besserer, J.~de Boer, J.~J.~Carroll,
%J.~Enders, T.~Hartmann, F.~K\"appeler, H.~Kaiser, U.~Kneissl,
%M.~Loewe, H.~J.~Maier, H.~Maser, P.~Mohr, P.~von Neumann-Cosel,
%A.~Nord, H.~H.~Pitz, A.~Richter, M.~Schumann, S.~Volz, A.~Zilges
\prl, 83, 5242
%
\bibitem[Dillmann et al.(2006)]{KADONISV0}
%KADoNiS - The Karlsruhe Astrophysical Database of Nucleosynthesis in Stars,
Dillmann, I., Heil, M., K\"appeler, F., Plag, R., Rauscher, T., \&
Thielemann, F.-K. 2006,
AIP Conf.\ Proc., 819, 123
%
\bibitem[Dillmann et al.(2009)]{KADONIS}
%KADoNiS v0.3,
% -- The third update of the "Karlsruhe
%Astrophysical Database of Nucleosynthesis in Stars",
Dillmann, I., Plag, R., K\"appeler, F., \& Rauscher, T. 2009,
Proc.\ of "EFNUDAT Fast Neutrons - scientific workshop on neutron
measurements, theory \& applications", April 28-30 2009, Geel, Belgium;
available at {\it{www.kadonis.org}}.
%
\bibitem[ENSDF(2010)]{ENSDF}
ENSDF 2010, ({\it{http://www.nndc.bnl.gov/ensdf/}}) online database; based on
\citet{Singh02} for $^{79}$Se, \citet{Basu10} for $^{95}$Zr, \citet{Ohya10} for
$^{121}$Sn, \citet{Bas10} for $^{187}$Os, and \citet{Ach06} for $^{193}$Pt.
%
\bibitem[Fowler(1974)]{fow74}
Fowler, W. A. 1974, \qjras, 15, 82
%
\bibitem[Frota-Pess\^{o}a \& Joffily(1986)]{Fro86}
  Frota-Pess\^{o}a, E. \& Joffily, S. 1986,
  Nuovo Cimento, 91A, 370
%
\bibitem[Fujii et al.(2010)]{Fuj10}
Fujii, K., et al.\ 2010,
\prc, 82, 015804
%
\bibitem[Gintautas et al.(2009)]{Gin09}
Gintautas, V., Champagne, A. E., Kondev, F. G., \& Longland, R. 2009,
\prc, 80, 015806
%
\bibitem[Heil et al.(2008)]{Heil08}
Heil, M., et al.\ 2008,
%M.\ Heil, N.\ Winckler, S.\ Dababneh, F.\ K{\"a}ppeler, K.\ Wisshak,
%S.\ Bisterzo, R.\ Gallino, A.\ M.\ Davis, T.\ Rauscher,
\apj, 673, 434
%
\bibitem[Hoffman et al.(1999)]{hoff99}
Hoffman, R. D., Woosley, S. E., Weaver, T. A., Rauscher, T., \& F.-K. Thielemann 1999, \apj, 521, 735
%
\bibitem[Holmes et al.(1976)]{holadndt}
Holmes, J. A., Woosley, S. E., Fowler, W. A., \& Zimmerman, B. A. 1976, At.\ Data Nucl.\ Data Tables, 18, 305
%
\bibitem[Huther et al.(2010)]{Huth10}
Huther, L., Loens, H. P., Mart\'inez-Pinedo, G., \& Langanke, K. 2010,
Europ.\ Phys.\ J.\ A, 47, 10

\bibitem[Iliadis(2007)]{iliadisbook} Iliadis, C. 2007, Nuclear Physics of Stars (Wiley, Weinheim)
%
\bibitem[K\"appeler et al.(2011)]{Kae11}
K\"appeler, F.,  Gallino, R., Bisterzo, S., \& Aoki, W. 2011,
Rev.\ Mod.\ Phys., 83, 157
%
\bibitem[K\"appeler \& Mengoni(2006)]{Kae06}
K\"appeler, F. \& Mengoni, A. 2006,
Nucl.\ Phys., A777, 291
%
\bibitem[Mohr et al.(2009)]{Mohr09}
Mohr, P., Bisterzo, S., Gallino, R., K\"appeler, F., Kneissl, U., \&
Winckler, N. 2009,
\prc, 79, 045804
%
\bibitem[Mohr, K\"appeler, \& Gallino(2007)]{Mohr07}
Mohr, P., K\"appeler, F., \& Gallino, R. 2007,
\prc, 75, 012802(R)
%
\bibitem[Mosconi et al.(2010a)]{Mos10a}
Mosconi, M., et al.\ 2010a,
\prc, 82, 015802
%
\bibitem[Mosconi et al.(2010b)]{Mos10b}
Mosconi, M., et al.\ 2010b,
\prc, 82, 015803
%
\bibitem[Ohya(2010)]{Ohya10}
Ohya, S. 2010,
Nucl.\ Data Sheets, 111, 1619
%
\bibitem[Rauscher(2006)]{raudeflect}
Rauscher, T. 2006, \prc, 73, 015804
%
\bibitem[Rauscher(2008)]{raugamma}
Rauscher, T. 2008, \prc, 78, 032801(R)
%
\bibitem[Rauscher(2010)]{energywindows}
Rauscher, T. 2010, \prc, 81, 045807

\bibitem[Rauscher(2011)]{tommyreview}
Rauscher, T. 2011, Int.\ J. Mod.\ Phys.\ E, 20, 1071
%
\bibitem[Rauscher et al.(2002)]{rhhw02}
Rauscher, T., Heger, A., Hoffman, R. D. \& Woosley, S. E. 2002, \apj, 576, 323
%
\bibitem[Rauscher \& Thielemann(2000)]{Rau00}
Rauscher, T. \& Thielemann, F.-K. 2000,
At.\ Data Nucl.\ Data Tables, 75, 1

\bibitem[Rauscher, Thielemann, \& Kratz(1997)]{rtk97}
Rauscher, T., Thielemann, F.-K., \& Kratz, K.-L. 1997, \prc, 56, 1613

\bibitem[Shizuma et al.(2005)]{Shi05}
Shizuma, T., et al.\ 2005,
\prc, 72, 025808
%
\bibitem[Singh(2002)]{Singh02}
Singh, B. 2002,
Nucl.\ Data Sheets, 96, 1
%
\bibitem[Sonnabend et al.(2003a)]{sonn03branch}
Sonnabend, K., et al.\ 2003a, \apj, 583, 506

\bibitem[Sonnabend et al.(2003b)]{Sonn03}
Sonnabend, K., Mohr, P., Zilges, A., Hertenberger, R., Wirth, H.-F., Graw, G., \&
Faestermann, T. 2003b,
\prc, 68, 048802
%
\bibitem[Straniero, Gallino, \& Cristallo(2006)]{Str06}
Straniero, O., Gallino, R., \& Cristallo, S. 2006,
Nucl.\ Phys., A777, 311
%
\bibitem[Takahashi \& Yokoi(1987)]{Tak87}
Takahashi, K. \& Yokoi, K. 1987,
At.\ Data Nucl.\ Data Tables, 36, 375
%
\bibitem[Ward \& Fowler(1980)]{WF80}
Ward, R. A. \& Fowler, W. A. 1980, \apj, 238, 266
%
\bibitem[Woosley et al.(1976)]{wooadndt}
Woosley, S. E., Fowler, W. A., Holmes, J. A., \& Zimmerman, B. A. 1976, At.\ Data Nucl.\ Data Tables, 22, 371
%
\bibitem[Woosley \& Howard(1978)]{woohow}
Woosley, S. E. \& Howard, W. M. 1978, \apjs, 36, 285
%
%
%
%\bibitem{Fro06}
%C.\ Fr\"ohlich {\it et al.},
%\prl {\bf 96}, 142502 92006).
%
%\bibitem{Wan11}
%S.\ Wanajo, H.-T.\ Janka, S.\ Kubono,
%\apj {\bf 729}, 46 (2011).
%
\end{thebibliography}
\end{document}